\documentclass[twocolumn,prd,nofootinbib,aps,floats,floatfix,amsmath,amssymb,secnumarabic,preprintnumbers]{revtex4} %
\usepackage[final]{graphicx}
\usepackage{amsmath}
\usepackage{bbm}
\usepackage{amsfonts}
\usepackage{amssymb}
\usepackage{latexsym}
\usepackage{graphicx}
\usepackage[english]{babel}
\usepackage{multirow}
\usepackage{float}
\usepackage{url}
\usepackage{hyperref}
\usepackage{slashed}
\usepackage{xcolor}
\usepackage{slashbox}

\newcommand{\be}{\begin{equation}}
\newcommand{\ee}{\end{equation}}
\newcommand{\ba}{\begin{array}}
\newcommand{\ea}{\end{array}}
\newcommand{\bea}{\begin{eqnarray}}
\newcommand{\eea}{\end{eqnarray}}
\newcommand{\sss}{\scriptscriptstyle}

\def\ket#1{\left| #1\right\rangle}
\def\nn{\nonumber}

\begin{document}
\preprint{UdeM-GPP-TH-15-244, UMISS-HEP-2015-03}
\title{Quark-flavored scalar dark matter}
\author{Bhubanjyoti Bhattacharya\footnote{bhujyo@lps.umontreal.ca}}
\affiliation{Physique des Particules, Universit\'e de Montr\'eal,
C.P. 6128, succ.\ centre-ville, Montr\'eal, QC, Canada H3C 3J7}
\author{James M.\ Cline\footnote{jcline@physics.mcgill.ca}}
\affiliation{Department of Physics, McGill University,
3600 Rue University, Montr\'eal, Qu\'ebec, Canada H3A 2T8}
\affiliation{Niels Bohr International Academy, University of Copenhagen, 
Blegdamsvej 17, DK-2100, Copenhagen, Denmark}
\author{Alakabha Datta\footnote{datta@phy.olemiss.edu}}
\affiliation{
Department of Physics and Astronomy,
University of Mississippi,
Lewis Hall, University, Mississippi, 38677 USA}
\author{Grace Dupuis\footnote{dupuisg@physics.mcgill.ca}}
\affiliation{Department of Physics, McGill University,
3600 Rue University, Montr\'eal, Qu\'ebec, Canada H3A 2T8}
\author{David London\footnote{london@lps.umontreal.ca}}
\affiliation{Physique des Particules, Universit\'e de Montr\'eal,
C.P. 6128, succ.\ centre-ville, Montr\'eal, QC, Canada H3C 3J7}

\begin{abstract}

It is an intriguing possibility that dark matter (DM) could have
flavor quantum numbers like the quarks.  We propose and investigate a
class of UV-complete models of this kind, in which the dark matter is
in a scalar triplet of an SU(3) flavor symmetry, and interacts with
quarks via a colored flavor-singlet fermionic mediator.  Such
mediators could be discovered at the LHC if their masses are $\sim 1$
TeV.  We constrain the DM-mediator couplings using relic abundance,
direct detection, and flavor-changing neutral-current considerations.
We find that, for reasonable values of its couplings, scalar flavored
DM can contribute significantly to the real and imaginary parts of the
$B_s$-$\bar B_s$ mixing amplitude.  We further assess the potential
for such models to explain the galactic center GeV gamma-ray excess.
\end{abstract}
\maketitle

\section{Introduction}

In recent years, model builders have entertained the idea that dark
matter (DM) comes in three generations like the matter particles of
the standard model (SM), and that its interactions with the SM are
governed by an approximate flavor symmetry.  In the design of such a
model, one must decide whether the dark matter carries quark or lepton
flavor.  In this paper we focus on quark-flavored dark matter, which
has previously been studied in
refs.\ \cite{Kile:2011mn}-\cite{Bishara:2015mha}.  A common element of
such models is the presence of an additional new particle, the
mediator that carries the quantum numbers of the standard model
quarks, to which the dark matter couples.

One must also decide whether the dark matter is a fermion or a scalar
(implying the opposite choice for the mediator).  So far, previous
studies have assumed the former, which we refer to as FDM.  The scalar
case, which we call SFDM, has some distinctive features that deserve
investigation; we aim to fill this gap in the present paper.  One
difference is that the colored fermionic mediators $\chi$ have a
larger production cross section at the Large Hadron Collider,
improving the prospects for their discovery or tightening constraints
on their masses relative to scalar mediators.

Another difference is that scalar DM $\phi$ can couple to the Higgs by
the renormalizable operator $\lambda|\phi|^2 |H|^2$ that leads to
Higgs portal interactions.  We will show that this naturally dominates
over the mediator interactions for setting the relic abundance and
indirect signals for light dark matter, putting SFDM on a similar
footing to minimal scalar dark matter in these respects.  However, for
heavy DM with mass $m_\phi \sim 450$-1000 GeV, mediator exchange with
annihilation to $t\bar t$ can dominate over Higgs portal
annihilations.  Moreover, the mediator exchanges can lead to important
effects for direct detection and flavor-changing neutral-current
(FCNC) processes.  

A further motivation for our study arises from indications of an
excess of multi-GeV energy gamma rays from the galactic center (GC),
whose origin is not obviously tied to known astrophysical sources
\cite{Hooper:2010mq,Hooper:2011ti,
  Abazajian:2012pn,Zhou:2014lva,Daylan:2014rsa,Calore:2014xka}.  There
has been considerable interest in dark matter annihilations into
standard model particles as a possible explanation of the signal,
including FDM models \cite{Agrawal:2014una}. Here we update the status
of scalar dark matter annihilations through the Higgs portal to fit
the GC excess, taking account of newer data sets provided by
refs.\ \cite{Calore:2014xka,Murgia}.

In the following we define the models (section \ref{models}), derive
constraints on the mediator masses/couplings ($m_\chi$ and
$\Lambda_{ij}$) from the LHC (section \ref{LHC_constraints}), and show
the implications for the couplings from requiring a thermal origin for
the abundance (section \ref{abundance}).  Constraints from indirect
detection, as well as the tentative evidence for the GC excess, are
examined in section \ref{Indirect_detection}, followed by a study of
direct detection (section \ref{Direct_detection}).  Additional bounds
on $m_\chi$ and $\Lambda_{ij}$ from FCNC searches are presented in
section \ref{FCNC_effects}.  In section \ref{benchmarks} we illustrate
the range of possible effects in this model
A summary of our findings is given in the
concluding section \ref{conclusions}.

\section{Models}
\label{models}

The largest quark flavor symmetry group is a product of three SU(3)'s,
SU(3)$_Q\times$SU(3)$_u\times$SU(3)$_d$, where $Q$ denotes quark
doublets and $u,d$ the weak singlets. If we took the dark matter
triplet to transform under one of these SU(3)'s, it would be natural
to invoke minimal flavor violation (MFV) \cite{D'Ambrosio:2002ex} to
suppress FCNCs in our model.  However, this transformation choice is
not necessary; it is more general to assume that the DM transforms
under its own SU(3)$_\phi$ group \cite{Agrawal:2014aoa}, which like
the others gets spontaneously broken by the mechanism that generates
the Yukawa couplings.  We adopt this more general approach here.

This leads to three possible models, characterized by the quantum
numbers of the mediator particle $\chi$.  All of them have
interactions of the form
\be
	\Lambda_{ij}\,\phi^*_i\, \bar\chi P_{L,R} q_j + {\rm h.c.} ~,
\label{couplings}
\ee
where $q_j$ stands for quark doublet $Q_j$ or singlets $u_j$, $d_j$,
and $P_{L,R}$ projects onto left- or right-handed states (left for
$Q_j$ and right for $u_j$, $d_j$).  We will denote the models by
$Q,u,d$, according to the kind of quarks which appear in
(\ref{couplings}), and which the mediator must resemble in most
respects.  The differences are that the mediator has no generation
index, and it is vector-like, having a mass $m_\chi \gtrsim 1\,$ TeV
(see sect.\ \ref{LHC_constraints} below) that is independent from
electroweak symmetry breaking.

In addition to the interactions with quarks, scalar dark matter can
couple to the Higgs via
\be
 \lambda_{ij}\, \phi^*_i\phi_j\,|H|^2 ~,
\label{higgs_coupling}
\ee
where $\lambda_{ij}$ is Hermitian.  At scales above that where flavor
symmetry is broken, one expects the flavor-symmetric form
$\lambda_{ij} = \lambda_0\,\delta_{ij}$, but this gets flavor-breaking
radiative corrections that we will discuss below.

\begin{figure}[h]
\centerline{
\includegraphics[width=1\columnwidth]{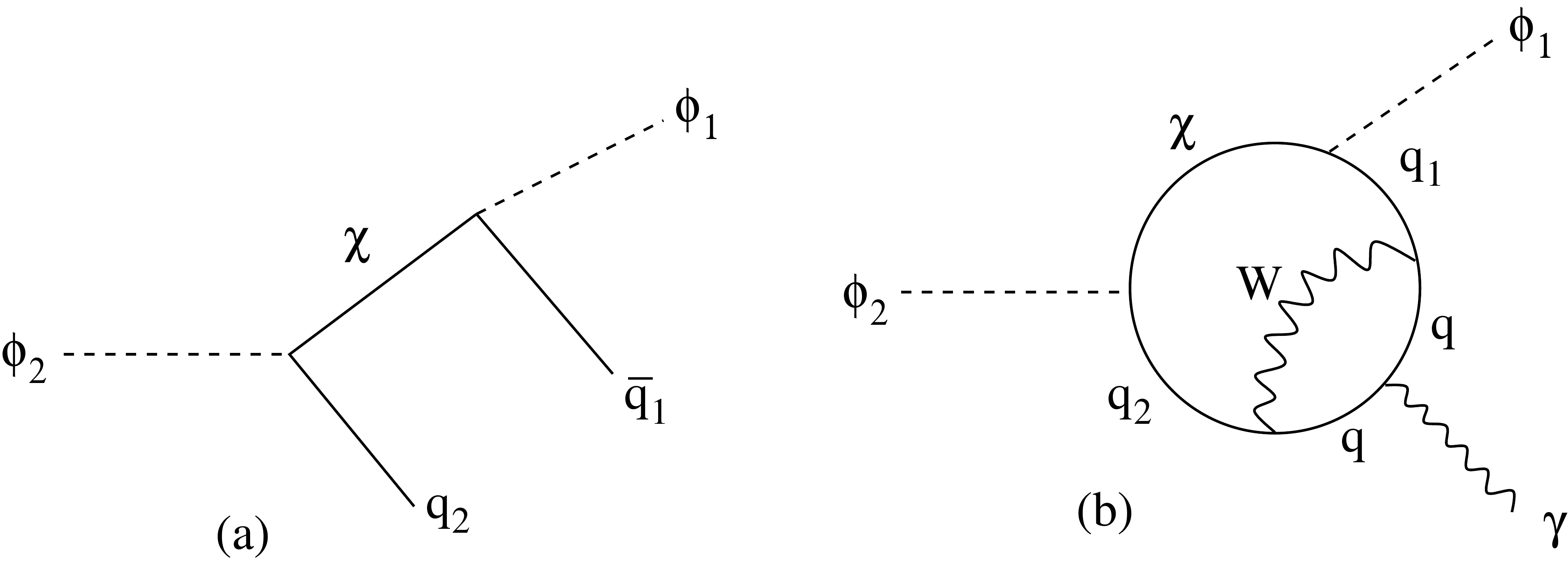}
}
\caption{Decay of the heavy DM state to the lightest one: (a)
  tree-level diagram for $\phi_2\to q_2\bar q_1 \phi_1$; (b) two-loop
  diagram for $\phi_2\to\phi_1\gamma$.}
\label{fig:dm-decay}
\end{figure}

\subsection{DM mass spectrum and couplings}

Like the coupling (\ref{higgs_coupling}), we expect the mass matrix
for scalar triplet dark matter to be flavor-conserving at high scales,
but corrected by flavor-breaking self-energies at one loop, and also
by the contribution from (\ref{higgs_coupling}) due to electroweak
symmetry breaking:
\be
	(m^2_{\phi})_{ij} = m^2_0\,\delta_{ij}
	+ m^2_1\, (\Lambda\Lambda^\dagger)_{ij} + v^2 \lambda_{ij}
	+ \dots ~,
\ee
where $v=174$ GeV is the VEV of the complex Higgs field.  {\it A
  priori,} there are no restrictions on the structure of
$\Lambda_{ij}$ nor do we know whether $m_0^2$ dominates over the other
contributions.  
%If $m_1^2|\Lambda\Lambda^\dagger|\gg v^2 |\lambda|$
%then the mass matrix can be approximately diagonalized by a unitary
%transformation $\phi \to U\phi$ such that $U^\dagger
%(\Lambda\Lambda^{\dagger})\,U$ is diagonal.  In that case, the
%interaction (\ref{couplings}) takes the form $\phi^\dagger\, {\rm
%  diag}(\Lambda) V\, \bar\chi P_{R,L} q$, where $V$ is also unitary,
%analogous to the CKM matrix, being the mismatch between $U$ and the
%transformation on $q$ needed to diagonlize its mass term.}  
For
simplicity of notation we will henceforth take $\Lambda_{ij}$ to
denote the matrix of couplings {\it in the DM/quark mass-eigenstate basis},
and allow the spectrum of DM states to be arbitrary, with $\phi_1$
being the lightest.

\subsection{Decays of excited DM states}

{\it A priori} we have three dark matter particles since $\phi_i$ is a
triplet.  As long as the mediators are heavier than the DM, the decay
$\phi\to \chi q$ is forbidden and the $Z_2$ symmetry under which both
$\phi$ and $\chi$ are charged guarantees the stability of $\phi_1$.
However, if there are mass splittings, as we generically expect there
to be, then only the lightest state is stable, since a heavier one
$\phi_2$ can decay via $\phi_2 \to q_2 \bar q_1 \phi_1$.  Even if the
mass splitting is too small to produce the quarks, they can be virtual
in a two-loop diagram to give $\phi_2 \to \phi_1\gamma$, as shown in
fig.\ \ref{fig:dm-decay}.  (In fact the photon must be off-shell since
the effective operator $\partial_\mu\phi^*_2 \partial_\nu\phi_1
F^{\mu\nu}$ vanishes for on-shell photons, but we can have for example
$\phi_2 \to \phi_1 e^+e^-$.)

\section{LHC constraints on mediators}
\label{LHC_constraints}

The colored fermionic mediators of the model may be produced at the
LHC, giving constraints on the mass of the new particle.  Pair
production of the mediator, with subsequent decay $\chi \rightarrow q
\phi$, contributes to a signal characterized by final-state jets and
missing transverse energy, denoted $\slashed{E}_{T}$. This is also the
signature of squark and gluino production in the supersymmetric
extension of the SM (SUSY). A recent ATLAS search for squarks and
gluinos in this final state was presented in ref. \cite{Aad:2014wea}.

Signals corresponding to different jet multiplicities are sensitive to
the production of mediators that couple either to tops or to light
quarks.  A mediator that couples to light quarks has an identical
signature to the light squark in SUSY, namely two jets and
$\slashed{E}_{T}$.  However, a colored fermion has a larger production
cross section than a scalar. The signature of a $t$-coupled mediator
is more similar to that of gluino production, having a final state
with higher jet multiplicity; the decays of $t$ and $\bar t$ in the
all-hadronic channel result in up to six jets.\footnote{Events with
  leptonic and semi-leptonic top-antitop decays are rejected in the
  analyses}

For this analysis, we simplify the model by ignoring the distinction
between DM flavors (valid as long as their masses are much less than
$m_\chi$), and we allow for two mediators $\chi_u$ and $\chi_d$.
These can represent either the two components of the SU(2)-doublet
$\chi$ in the $Q$ model, or the SU(2)-singlet mediators of the $u$ or
$d$ models.  The interaction terms can then be written as
\begin{equation}
\lambda_{u_{i}}\, \phi\, \bar{\chi}_{u} P_{L(R)} u_{i} + 
\lambda_{d_i}\, \phi\, \bar{\chi}_{d} P_{L(R)} d_{i} + h.c. 
\end{equation}
We used \textsc{MadGraph5} \cite{Alwall:2014hca} to calculate signal
cross sections and to generate parton-level events. Implementation of
the model in \textsc{MadGraph} is achieved with \textsc{FeynRules}
\cite{Alloul:2013bka}.

\begin{figure*}
\centering
\includegraphics[width=\columnwidth]{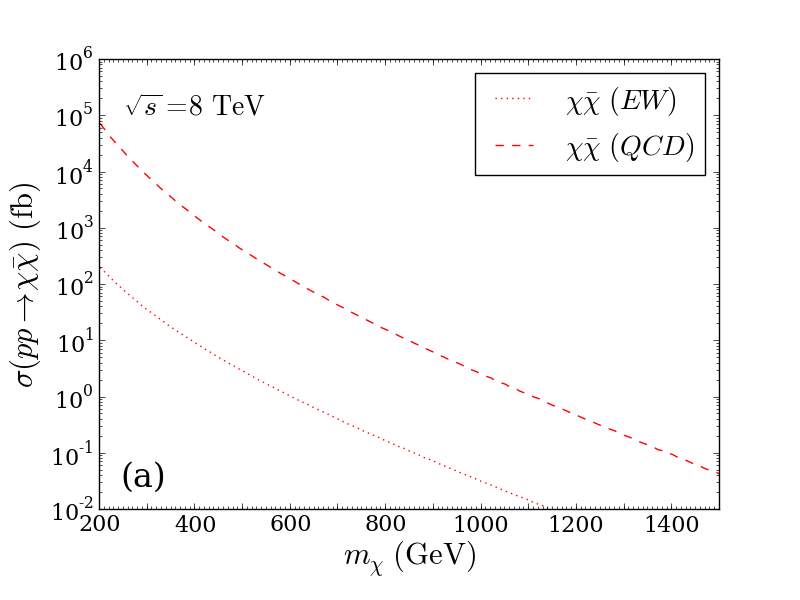}
\includegraphics[width=1\columnwidth]{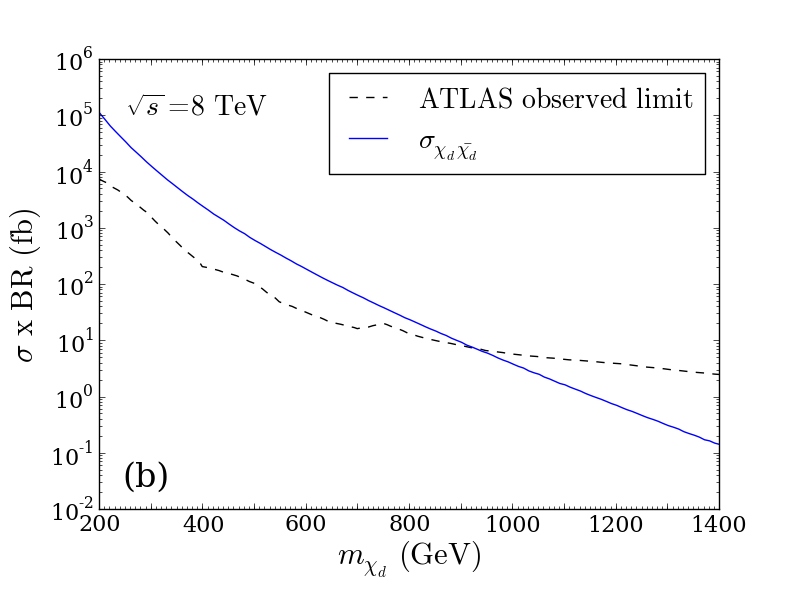}
\caption{Left (a): comparison of the relative magnitudes of electroweak
  and QCD contributions to mediator production cross section at the
  LHC for c.m.\ energy $\sqrt{s}=8$ TeV.  Right (b): lower bound on
  the mass of a colored mediator coupling to light quarks, resulting
  from the upper limit on $\chi \bar{\chi}$ production cross section in
  final states with jets and $\slashed{E}_{T}$.}
\label{fig:xxprod}
\end{figure*}

\begin{figure*}
\centerline{
    \includegraphics[width=2\columnwidth]{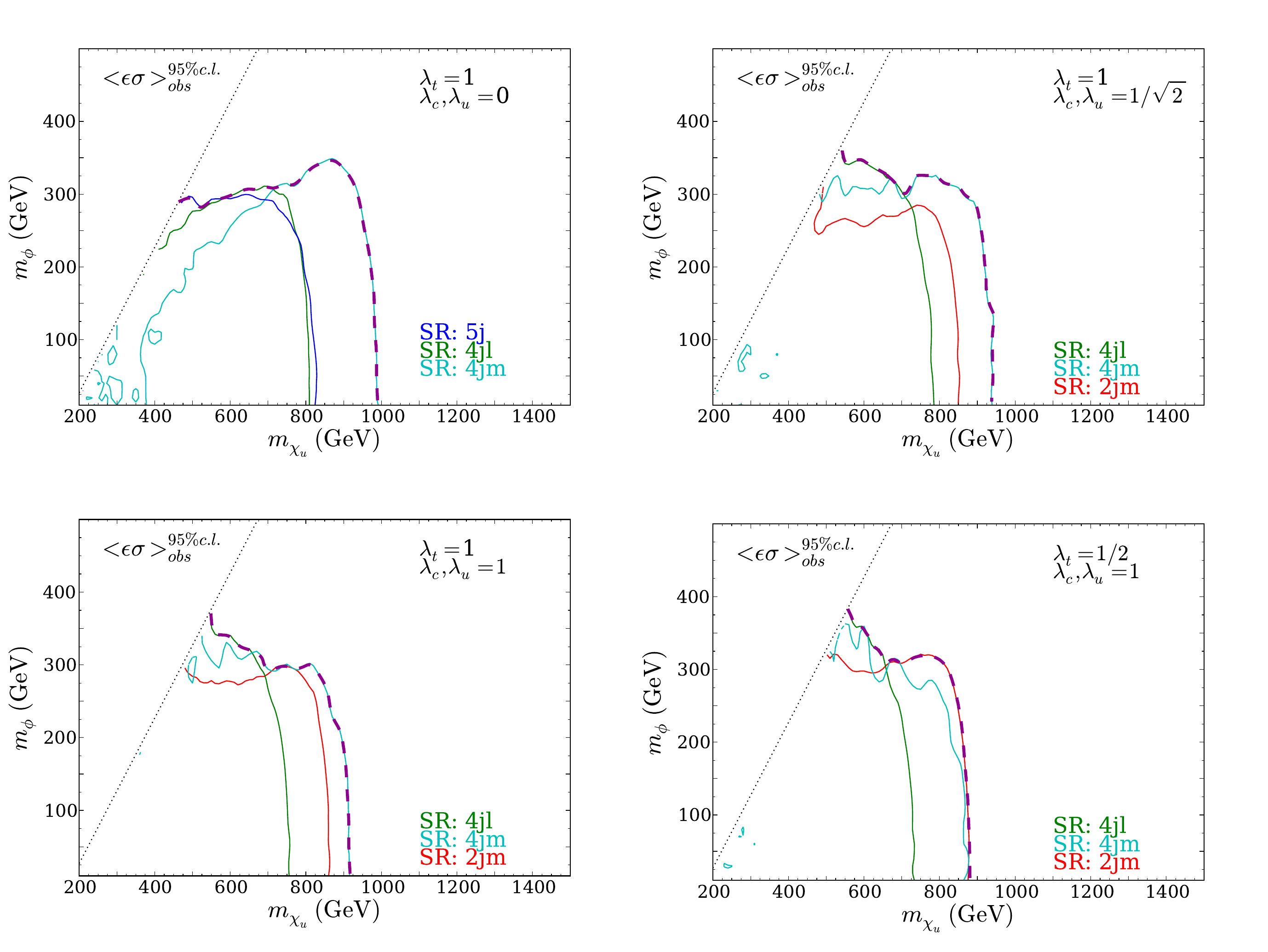}}
  \caption{ATLAS constraints on DM mass versus up-type mediator mass
    for different branching fractions of $\chi\to\phi\,t$ (as opposed
    to decays into light quarks): BF $ = 1$, $0.5$, $0.3$, and $0.1$
    from left to right and top to bottom.  Regions below and to the
    left of the dashed curves (envelope of exclusion from signal
    regions with 2, 4 and 5 jets) are excluded.}
\label{fig:umed_lims}
\end{figure*}

The electroweak contributions to the mediator production cross section
are highly subdominant to the QCD process. Fig.\ \ref{fig:xxprod}(a)
shows the leading-order (LO) cross sections for the two subprocesses,
verifying that the electroweak contributions may be neglected, as one
would expect. The limits on $\chi_u$ and $\chi_d$ are thus equally
applicable to the mediators of the $u$ and $d$ models, respectively.

We first consider the $\chi_d$ mediator.  Ref.\ \cite{Aad:2014wea}
provides an upper limit on the pair-production cross section for light
squarks as a function of their mass. Under the assumption that the
signal topology does not differ substantially for the mediator signal,
we calculate the cross section for $\chi_{d} \bar{\chi}_{d}$
production and translate this limit to a 95 \% c.l.\ bound on the
mediator mass, applying a $K$-factor to account for higher-order
corrections. The hadronic production mechanism of $\chi$ is the same
as for any colored fermion: we therefore estimate the $K$-factor to be
the same as for $t\bar{t}$ production, and obtain a value $K=1.5$,
comparing the NNLO value of the top pair-production cross section at
$\sqrt{s}=8$ TeV \cite{Aad:2015pga,Cacciari:2011hy} to the value
calculated at LO. The limit is shown in fig.\ \ref{fig:xxprod}(b). We
find that the mass of a down-coupled mediator is constrained to be
$\gtrsim 920$ GeV, regardless of the coupling strength, as long as
$\chi$ decays within the detector.

In the case of $\chi_u$, the event topology of the signal may be
substantially different than that from squark decays (other than
$\tilde t$) due to the possible decay channel $\chi_u \to \phi\, t$.
If this channel is suppressed, the signal is identical to that of
$\chi_d$, and the same limit $m_\chi \gtrsim 920$ GeV applies.  A
different approach is necessary for the $\chi_u \to \phi\, t$ channel.
In this case we use the ATLAS upper limit on the visible cross
section, defined as the product of (cross section) $\times$
(reconstruction efficiency) $\times$ (signal acceptance), in other
words, an effective cross section for the number of signal events
observed. To obtain a limit, we simulate full events with
hadronization and detector simulation in order to determine the signal
acceptance and reconstruction efficiency of the mediator signal.
Details are provided in appendix \ref{app}.

The resulting limit on $\chi_u \to \phi\, t$, obtained from the 95\%
c.l.\ upper bound on the visible signal cross section, is shown in
fig.\ \ref{fig:umed_lims}.  The exclusion region in the $(m_{\chi_u},
m_{\phi})$ plane is shown for the signal regions (SRs) with the
highest sensitivity, and thereby the greatest exclusion reach. These
correspond to SRs having four jets, with both loose and medium-level
kinematic cuts (4jl and 4jm), the five-jet signal region (5j) in the
case of 100\% decays to top quarks, and the two-jet signal region,
with medium-level cuts (2jm), for the other cases. For light DM,
$m_{\phi} \lesssim 400$ GeV, the limit corresponds to a lower bound on
a top-coupled mediator mass of $\sim 1$ TeV in the case of decays
exclusively to tops. The bound relaxes with branching fraction to
$\sim 900$ GeV; the exclusion by the 4j SR is relaxed, while that of
the 2j SR becomes stricter, as the branching fraction to $t\phi$ is
decreased.\footnote{As the third-generation coupling is taken to zero,
  the limiting value of the lower bound on $m_\chi$ is slightly lower
  than in the light quark case, fig.\ \ref{fig:xxprod}(b). We adopt
  the latter constraint, as the discrepancy is most likely a result of
  using different simulation and reconstruction methods than those of
  ref.\ \cite{Aad:2014wea}, as well as some subtler differences
  between the analyses.}

\begin{figure}[t]
\centerline{
\includegraphics[width=1\columnwidth]{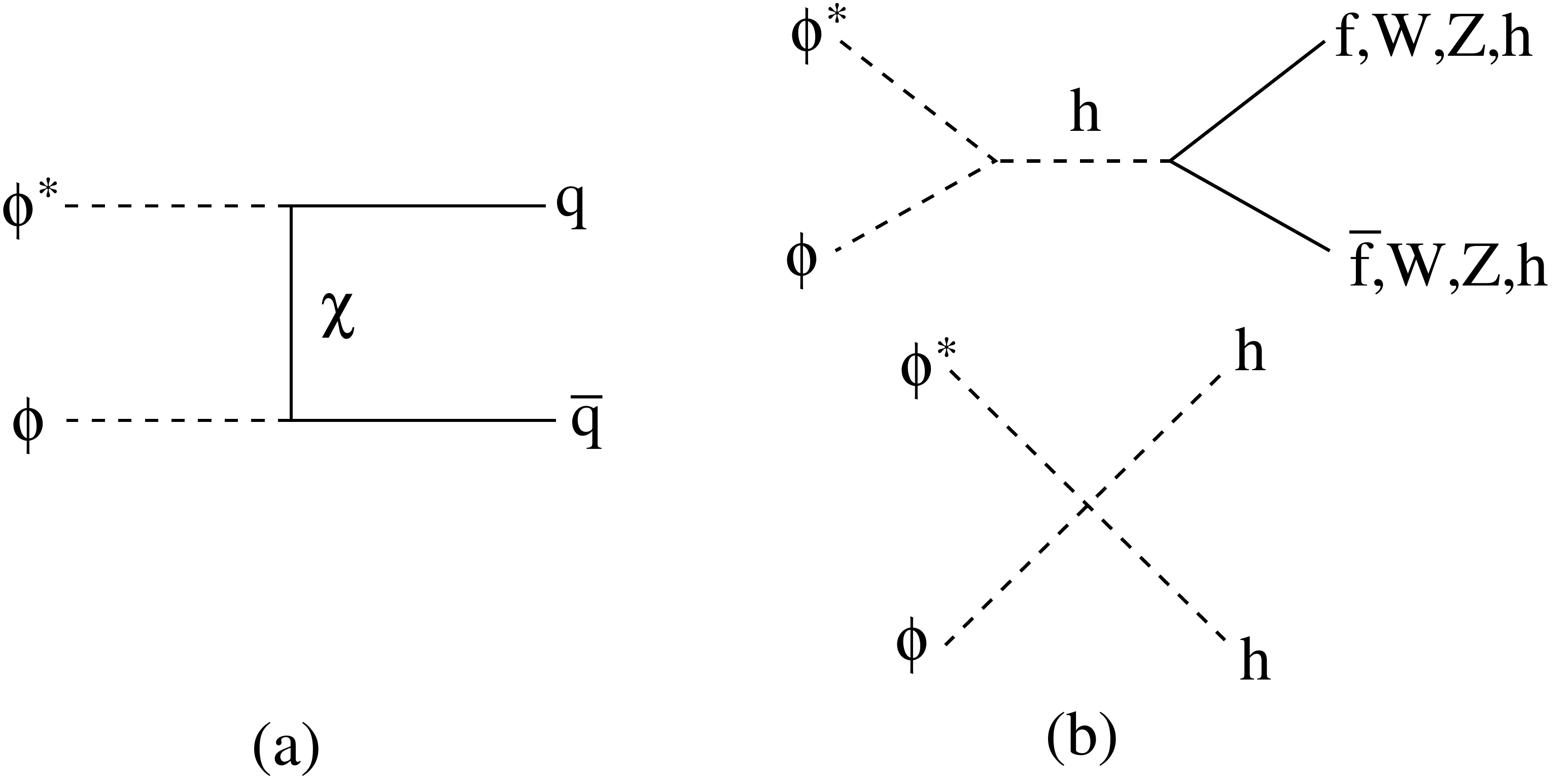}
}
\caption{Processes contributing to thermal freezeout of scalar dark
  matter: (a) by exchange of the mediator $\chi$; (b) through Higgs
  portal interactions.  Possible decays of $h$ in the lower diagram
  are not shown.}
\label{fig:relic}
\end{figure}

\section{Relic abundance}
\label{abundance}

Since our DM candidate is a complex scalar, its particle and
antiparticle are distinct and it could therefore be an example of
asymmetric dark matter, whose abundance arises through the generation
of a particle-antiparticle asymmetry in the early universe.  However,
this would require a more complicated model, so we will assume that
such an asymmetry is negligible and that the relic abundance comes
from thermal freezeout of the annihilation processes.  These can
proceed either through $t$-channel exchange of the mediator $\chi$ or
the $\lambda_{ij}\, \phi^*_i\phi_j|H|^2$ coupling, as shown in
fig.\ \ref{fig:relic}. It will turn out that the former is the
dominant process only in models with annihilation to top quarks, and
with $m_\phi$ exceeding some minimum value to be determined. We
consider $\chi$-mediated annihilations first, and subsequently treat
the Higgs portal scenario, constraining $\lambda_{11}$ as a function
of $m_{\phi_1}$ in order to get the observed abundance of dark matter.

\begin{figure*}[t]
\centerline{
\includegraphics[width=0.97\columnwidth]{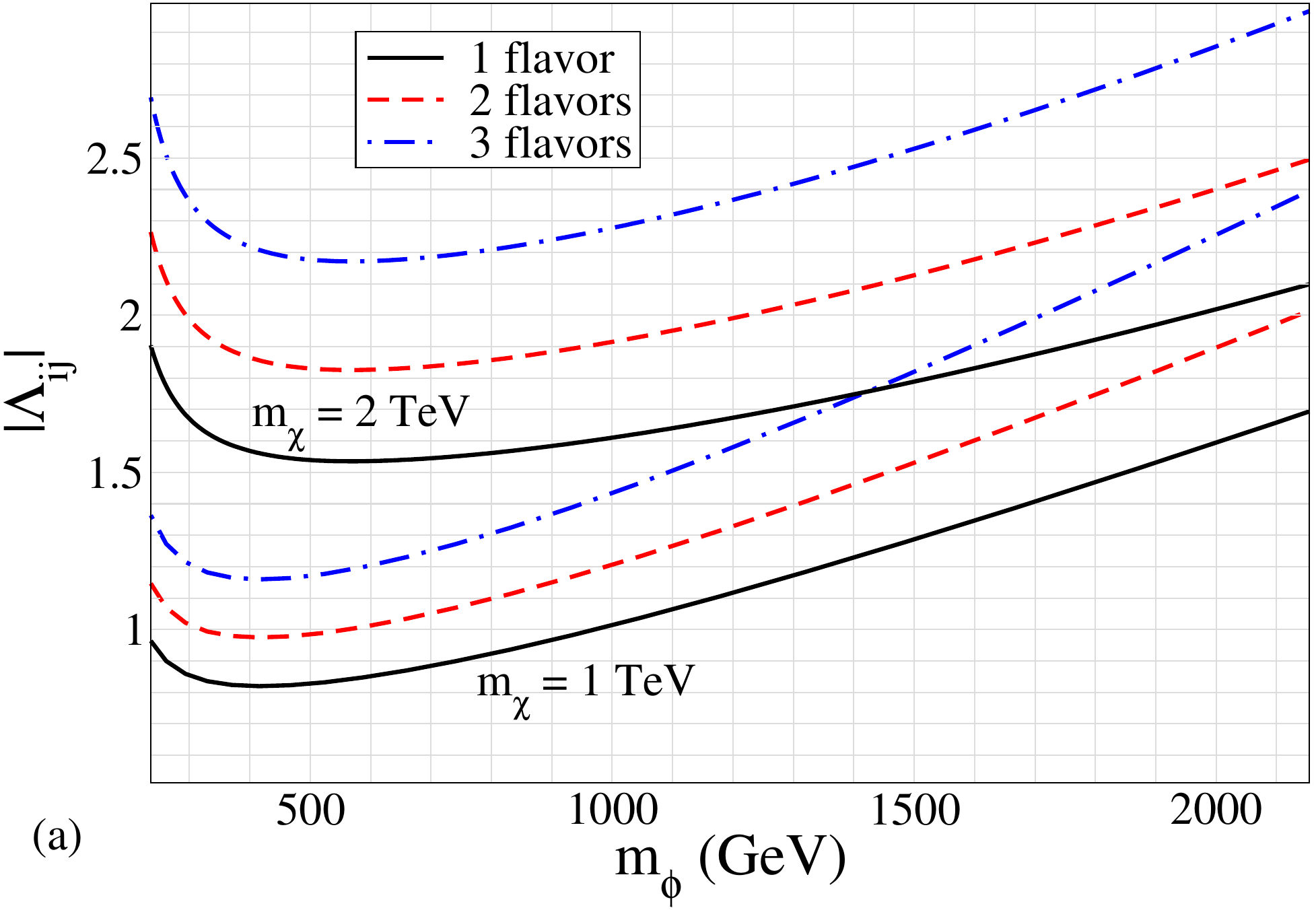}
\includegraphics[width=1\columnwidth]{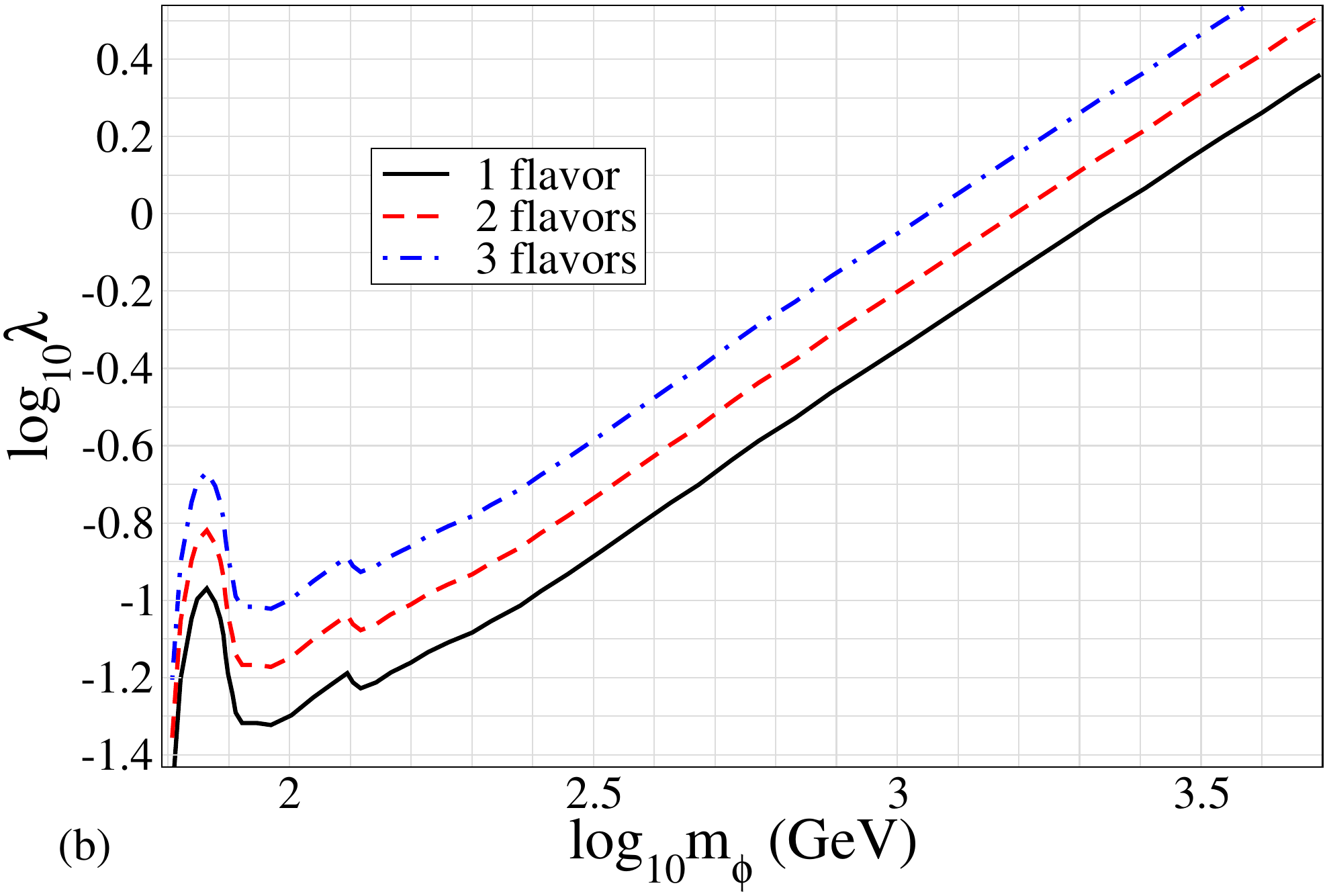}
}
\caption{Left (a): value of DM-mediator-quark coupling
  $|\Lambda_{ij}|$ needed for thermal relic abundance from
  annihilations via $t$-channel mediator exchange, as a function of DM
  mass $m_\phi$, assuming $\phi^*\phi\to t\bar t$ and $m_\chi = 1$ TeV
  (lower curves) or $m_\chi = 2$ TeV (upper curves).  Different curves
  show the dependence upon how many flavors of DM are in equilibrium
  at the time of freezeout. Right (b): value of DM-Higgs cross
  coupling needed for thermal relic abundance from Higgs portal
  annihilations, again showing the dependence on number of DM flavors
  in equilibrium at freezeout.}
\label{fig:lambda-relic}
\end{figure*}

\subsection{Mediator dominance}
\label{med_dom}

We begin by evaluating the amplitude in fig.\ \ref{fig:relic}(a) for
$\phi^*_i\phi_k \to q_l\bar q_j$.  In general, the final-state quarks
could be different from each other, and likewise the initial dark
matter flavors could be distinct.  To simplify the kinematics we will
evaluate the cross section in the approximation that the DM mass
splittings and quark masses are small compared to the average
$m_\phi$.  At kinematic threshold, where the DM is at rest, the
spin-summed, squared matrix element is
\be
	|{\cal M}|^2 = 6|\Lambda_{ij}\Lambda^*_{kl}|^2\,
	{ m_q^2\,(m_\phi^2 - m_q^2)\over (m_\phi^2 + m_\chi^2
	-m_q^2)^2} ~,
\label{med-matrix-element}
\ee
including the sum over colors.  The annihilation cross section is then
given by
\be
	\sigma v_{\rm rel} = {|{\cal M}|^2\over 32\pi m_\phi^2}
	\sqrt{1-(m_q/m_\phi)^2} ~.
\ee
To get the right relic density, we can match this to the value found
in ref.\ \cite{Steigman:2012nb}, where the required cross section as a
function of mass is derived.  More specifically, for complex scalar
DM, the required cross section is twice as large as that given in
\cite{Steigman:2012nb}, where self-conjugate DM was assumed.
Moreover, if the higher-mass DM states are in thermal equilibrium at
the time of freeze-out, we must multiply the fiducial value of the
cross section for a single complex scalar by the total number of
complex DM components.

The result is shown in fig.\ \ref{fig:lambda-relic}(a), assuming a
mediator mass of $m_\chi = 1$ TeV, and considering only the case where
the final state is $t\bar t$, since for the lighter quarks, the
suppression by $m_q^2$ leads to nonperturbatively large values of
$\Lambda_{ij}$.  Thus mediator exchange can only be the dominant
contribution to annihilation in the $Q$ and $u$ models.

\subsection{Higgs portal dominance}

In the case where the Higgs portal interactions dominate the dark
matter annihilation cross section, the required value of
$\lambda_{ij}$ can be deduced by rescaling with respect to real scalar
singlet dark matter, which has been studied in detail in many
references, including \cite{Cline:2013gha}.  Since the abundance
scales as $1/\langle\sigma v\rangle\sim 1/\lambda_{ij}^2$, we must
increase $\lambda_{ij}$ by a factor of $\sqrt{2}$ for complex $\phi$
relative to a real singlet, to compensate for the doubling of the
number of degrees of freedom.  In our model, there are actually three
complex scalar dark matter states, because of the flavor multiplicity.
If they are all degenerate, then $\lambda_{ij}$ must be increased by a
further factor of $\sqrt{2}^2$ relative to the complex singlet case.
The exact value required will depend upon the mass splittings of the
DM matter states and the thermal history.  In particular, if the
heavier DM states decay before freezeout, they will not contribute to
the final abundance, whereas if they decay afterwards, they will.  The
range of possibilities is covered by the three curves shown in
fig.\ \ref{fig:lambda-relic}(b).

\begin{figure}[t]
\centerline{
\includegraphics[width=0.8\columnwidth]{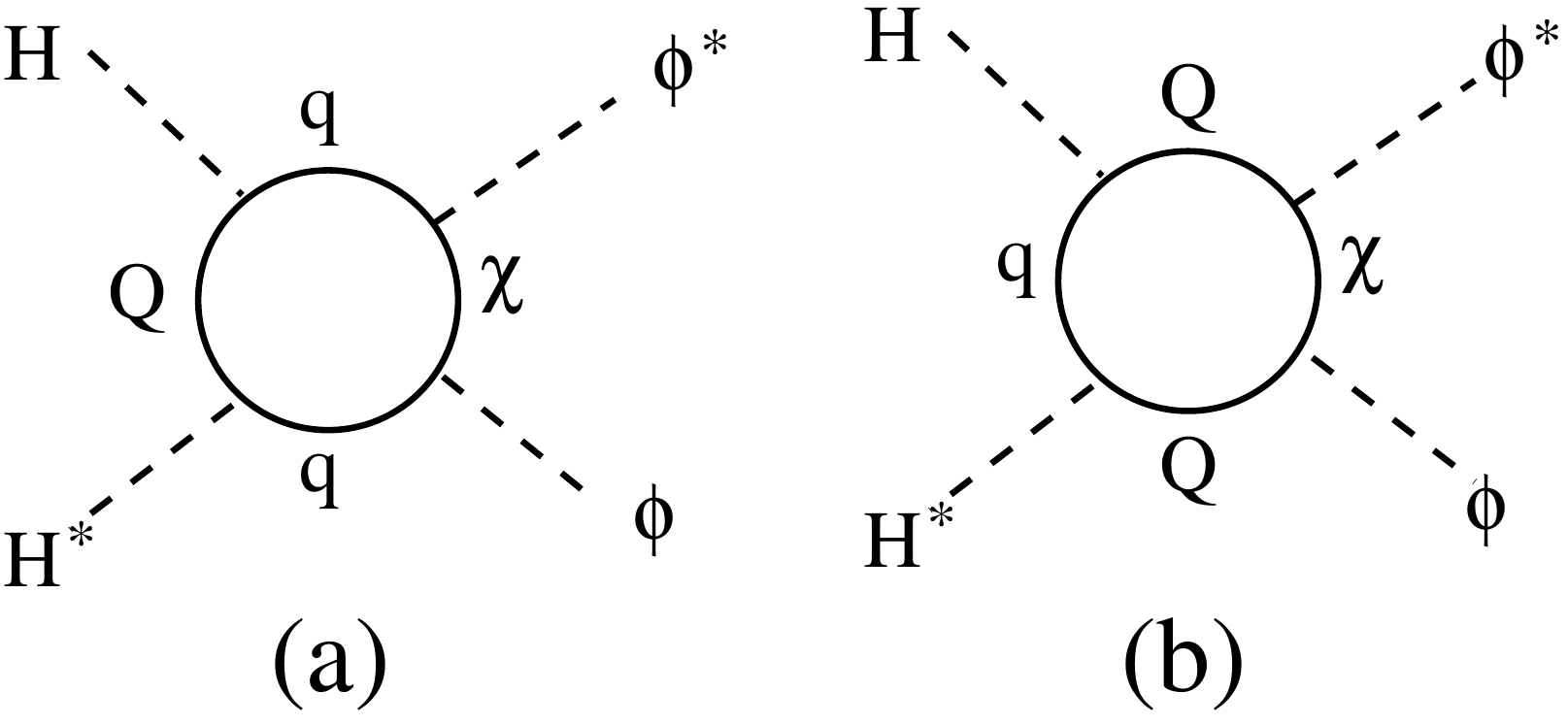}
}
\caption{One-loop contribution to $\lambda_{ij}\phi^*_i\phi_j|H|^2$
  interaction, where $q\, (Q)$ stands for electroweak singlet
  (doublet) quarks, and the routing of weak isospin is shown for (a)
  singlet and (b) doublet mediators, respectively.}
\label{fig:loop-fig}
\end{figure}

\begin{figure}[t]
\centerline{
\includegraphics[width=\columnwidth]{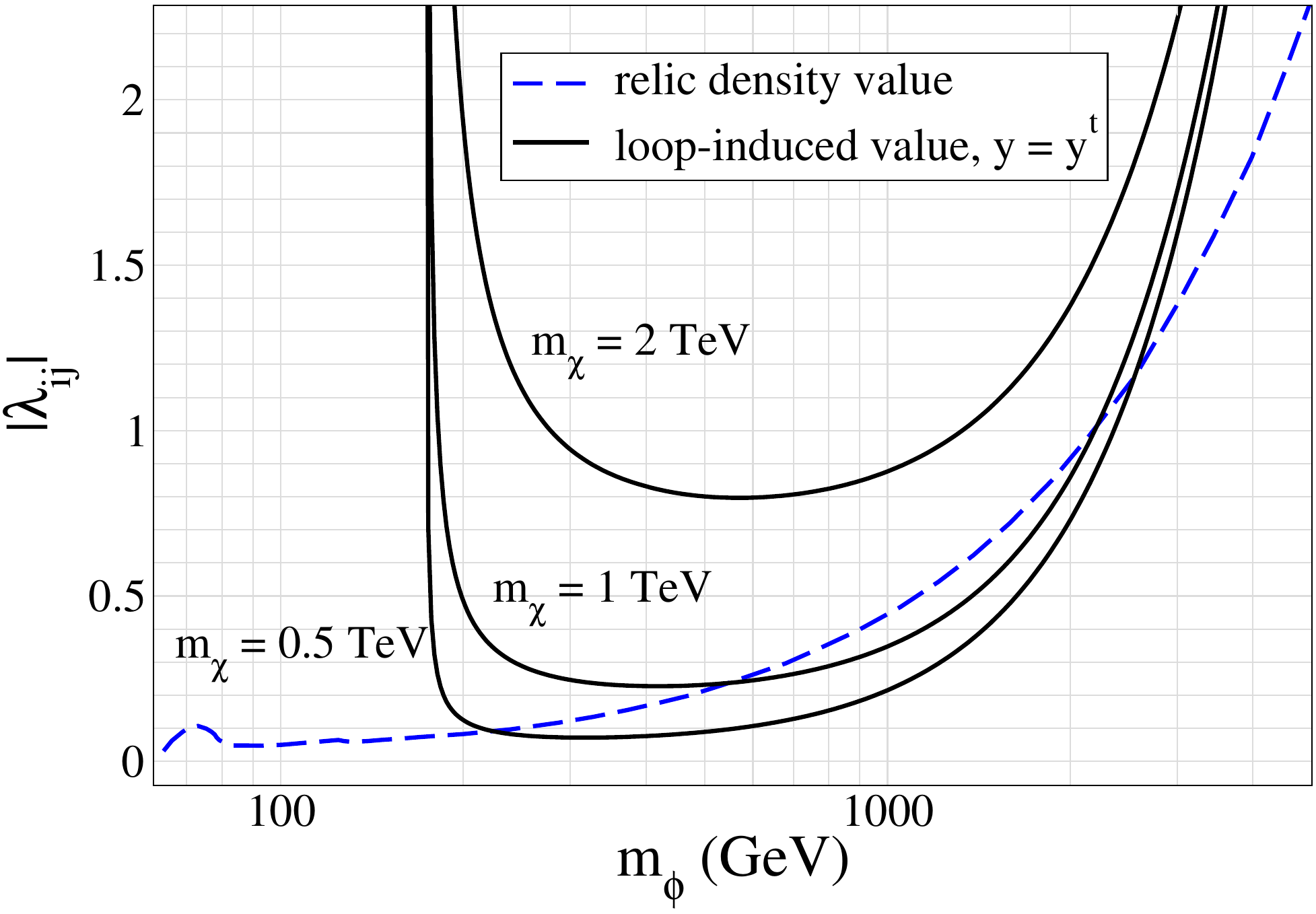}
}
\caption{Comparison of the value of $|\lambda_{ij}|$ needed for
  correct relic density via Higgs portal annihilations (dashed curve)
  with the value coming from loop contributions,
  eq.\ (\ref{loop-estimate}), in the case where $|\Lambda_{ij}|$ is
  large enough for mediator exchange to give the right relic density
  (also assuming that $y_u$ rather than $y_d$ is the Yukawa coupling
  matrix appearing in the loop).  Models where the solid curve is
  higher than the dashed one have annihilations dominated by the Higgs
  portal, in the absence of fine tuning.}
\label{fig:lambda-comp}
\end{figure}

\begin{figure*}[t]
\centerline{
\includegraphics[width=2\columnwidth]{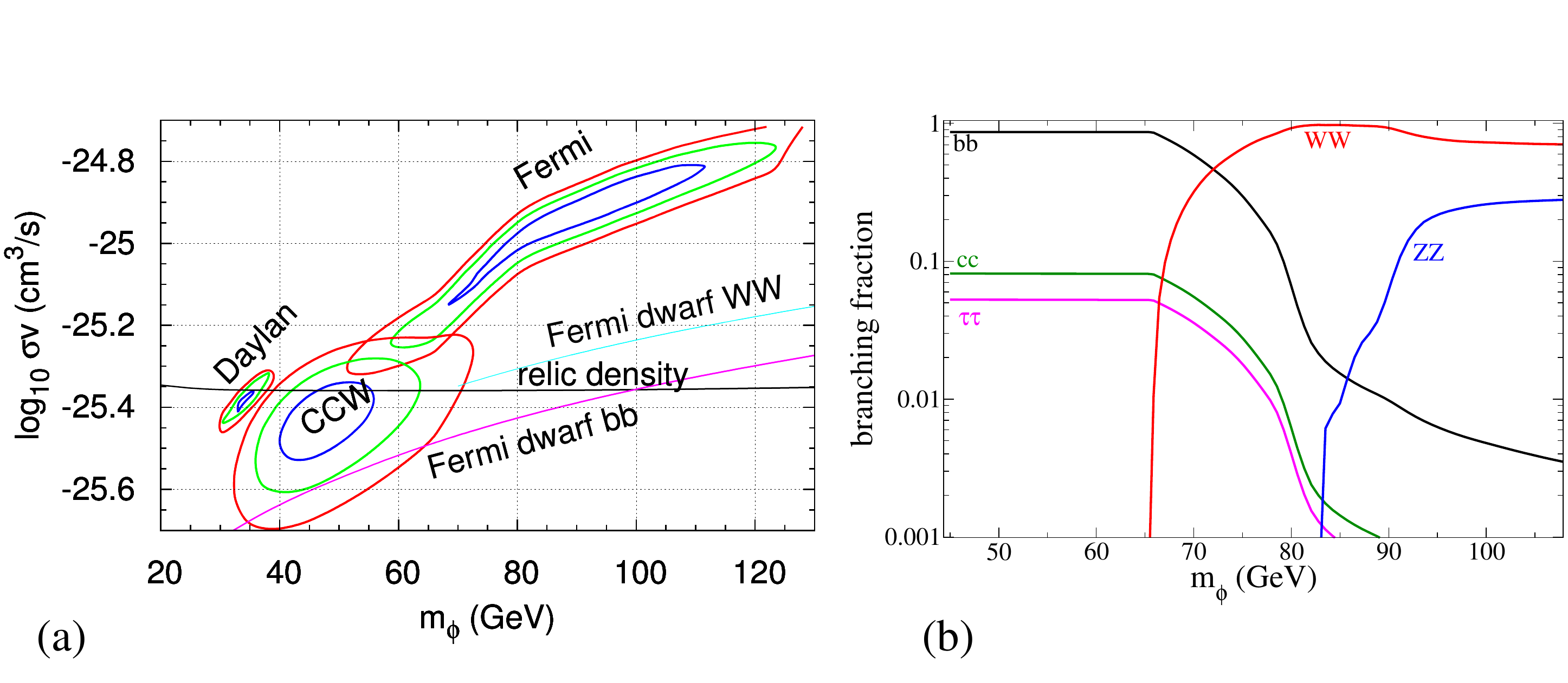}
}
\caption{Left (a): cross section for $\phi^*\phi$ to annihilate
  through the Higgs portal, for relic density, galactic center gamma
  ray excess (1, 2 and 3$\sigma$ contours for three data sets:
  Daylan {\it et al.,} CCW and Fermi), and Fermi/LAT upper limit from
  dwarf spheroidal galaxies assuming $b\bar b$ or $WW$ final states.
  Right (b): branching fractions for $\phi^*\phi$ to annihilate into
  SM final states through the Higgs portal (adapted from
  ref.\ \cite{Cline:2013gha}).  }
\label{fig:indirect}
\end{figure*}

\subsection{When can mediators dominate?}
\label{when}

In order to determine in which cases the Higgs portal interactions
dominate over mediator exchange for setting the relic density, we note
that the couplings $\lambda_{ij}$ can naturally be no smaller than
typical values generated by the loop diagrams shown in
fig.\ \ref{fig:loop-fig}.  One could fine tune the bare value of
$\lambda_{ij}$ to cancel the loop contribution, but in the absence of
such tuning one would expect a minimum magnitude for $\lambda_{ij}$,
which we estimate by taking the leading logarithmic contribution and
evaluating the log between the DM mass scale of order 100 GeV and a UV
scale $\Lambda = 10$ TeV, which we take to be the minimum scale of
validity for our model, considered as a low-energy effective theory.
In this case, $\ln(\Lambda^2/m_\phi^2) \cong 9$, and by evaluating the
loop we get the estimate
\be
	|\lambda_{ij}| \gtrsim {27\over 8\pi^2} (\Lambda y y^\dagger
	\Lambda^\dagger)_{ij}
\cong {27\,\Lambda_{i3}\Lambda^*_{j3}\over 8\pi^2 v^2}\left\{\begin{array}{ll}
	m_b^2,& y=y_d\\ m_t^2,& y = y_u\end{array}\right. ~,
\label{loop-estimate}
\ee
where $\Lambda_{ij}$ is the $\phi\bar\chi q$ coupling and $y_{ij}$ the
Yukawa coupling relevant to the particular model of interest; $v=174$
GeV is the complex Higgs VEV.  Which Yukawa matrix appears depends
upon the mediator.  If the mediator is $u$- or $d$-like, then $y =
y_u$ or $y_d$ respectively.  But if it is the doublet ($Q$-like), then
we must sum over both possibilities, in which case $y_u$ dominates.
In either case, working in the basis of diagonal Yukawa matrices gives
the approximation shown in (\ref{loop-estimate}).

By substituting the value of $|\Lambda_{ij}|$ shown in
fig.\ \ref{fig:lambda-relic}(a) into eq.\ (\ref{loop-estimate}), and
comparing to the value of $\lambda_{ij}$ shown in
fig.\ \ref{fig:lambda-relic}(b), we can determine when it would be
inconsistent to assume that mediator exchange dominates over Higgs
portal interactions.  This comparison is shown in
fig.\ \ref{fig:lambda-comp}, for models where $y_u$ rather than $y_d$
appears in the loop (otherwise, the solid curve is lower by a factor
of $(m_b/m_t)^2$, giving no useful constraint).  We assumed for these
curves that only one DM flavor is in equilibrium; for higher numbers,
both curves scale upward by the same factor, so that the values of
$m_\phi$ where they intersect do not change.

It is interesting to notice that the same model-building choices that
would suppress the loop contribution (\ref{loop-estimate}) also
suppress the mediator contribution to annihilation.  In particular,
the models for which $y = y_d$ in (\ref{loop-estimate}) are those
where the mediator is $d$-like, but these have cross sections for
$\phi\phi\to q\bar q$ suppressed at least by $m_b^2$ in
eq.\ (\ref{med-matrix-element}), making it impossible to satisfy the
relic density constraint with reasonable values of $\Lambda_{ij}$.

The upshot of this analysis is that only in the $Q$ and $u$ models
with $300 {\rm\ GeV} \lesssim m_\phi < m_\chi$  and $m_\chi\cong
0.5$-1 TeV can we consistently assume mediator dominance of the
annihilation cross section.  Here we have taken advantage of the fact
that our LHC constraint on $m_\chi$ is weaker for $m_\phi \sim 300$
GeV than for lighter $m_\phi$; see fig.\ \ref{fig:umed_lims} (upper
left).  Moreover, the tree-level value of $\lambda_{ij}$ can exceed
the minimum coming from the loop estimate in
eq.\ (\ref{loop-estimate}); thus Higgs portal dominance is always a
logical possibility, even when not a necessity.

\section{Indirect detection}
\label{Indirect_detection}

Annihilation of DM in our galaxy or neighboring ones can produce gamma
rays from the decays of final-state particles.  The Fermi Large Area
Telescope (LAT) continues to improve constraints on dark matter
annihilation into various final states, from observations of dwarf
spheroidal galaxies that are relatively uncontaminated by baryonic
background signals \cite{Ackermann:2015zua}.  The constraints are
strongest for light dark matter, whose relic density is higher. They
are therefore relevant in the region of parameter space where
annihilation is primarily through the Higgs portal.

For $m_\phi$ below the $W/Z$ threshold, annihilation is almost
exclusively into $b\bar b$.  We reproduce the Fermi limit from
ref.\ \cite{Ackermann:2015zua} on the annihilation cross section into
$b\bar b$ in fig.\ \ref{fig:indirect}(a), where it is relaxed by a
factor of 2 due to the dark matter not being self-conjugate in our
model.  The value needed for the observed relic density (also
increased by the factor of 2) is also shown, suggesting that masses
below 100 GeV are ruled out.  However, for $m_\phi > 70$ GeV, the
dominant annihilation channel is no longer $b\bar b$ but rather $WW$
or $WW^*$ where one of the $W$'s is off shell;
fig.\ \ref{fig:indirect}(b) shows the branching fractions into
different final states.  The constraint on the $WW+WW^*$ channel is
weaker by a factor 1.3, which can allow for somewhat lighter dark
matter ($m_\phi\sim 80$ GeV) to be consistent with both relic density
and indirect constraints.  The actual constraint on Higgs portal
models (not determined by ref.\ \cite{Ackermann:2015zua}) interpolates
between the $WW$ and $b\bar b$ curves in the region $m_{\phi} = 70$-80
GeV.

\subsection{Galactic center $\gamma$-ray excess}

A possible signal in Fermi/LAT data for dark matter annihilation in
the galactic center has been discussed by several groups, most
recently in refs.\ \cite{Daylan:2014rsa,Calore:2014xka,Murgia}
(referred to here as Daylan {\it et al.}, CCW and Fermi respectively).
The Fermi collaboration itself presented preliminary evidence for
gamma rays in excess of those attributable to known astrophysical
sources in the central $15^\circ\times 15^\circ$ region of the galaxy
\cite{Murgia,Porter:2015uaa}.  Recently, new evidence has been
presented in favor of unresolved millisecond pulsars as a likely
astrophysical source
\cite{Porter:2015uaa,Cholis:2015dea,Bartels:2015aea,Brandt:2015ula},
but pending a definite resolution, it is interesting to explore
whether dark matter models can consistently explain the observations.

Here we have used a similar methodology as in
ref.\ \cite{Cline:2015qha} to fit $m_\phi$ and its annihilation cross
section $\langle\sigma v\rangle$ to the excess signal as characterized
respectively by Daylan {\it et al.}, CCW and Fermi.  To generate the
predicted signal, we compute the photon spectrum from annihilation
into SM states with the branching ratios shown in
fig.\ \ref{fig:indirect}(b), using spectra from
ref.\ \cite{Cirelli:2010xx}.  These are compared to the data to
compute $\chi^2$ statistics for which the 1, 2 and 3$\sigma$-allowed
regions are shown in fig.\ \ref{fig:indirect}(a).

The three data sets are not fully consistent with one another, and
they conflict with the Fermi dwarf constraint except for part of the
CCW $3\sigma$ region.  This region however has too small an
annihilation cross section with respect to that needed for the relic
density, by a factor of $\cong 1.2$, which would lead to a 20\%
increase in the dark matter abundance.  The experimental error in the
observed abundance as determined by Planck is about 4\%.  A consistent
interpretation would require that the actual excess signal be somewhat
lower in intensity, as may be the case if part of it is due to
millisecond pulsars.

\section{Direct detection}
\label{Direct_detection}

\begin{figure}[b]
\begin{center}
\includegraphics[width=0.4\textwidth]{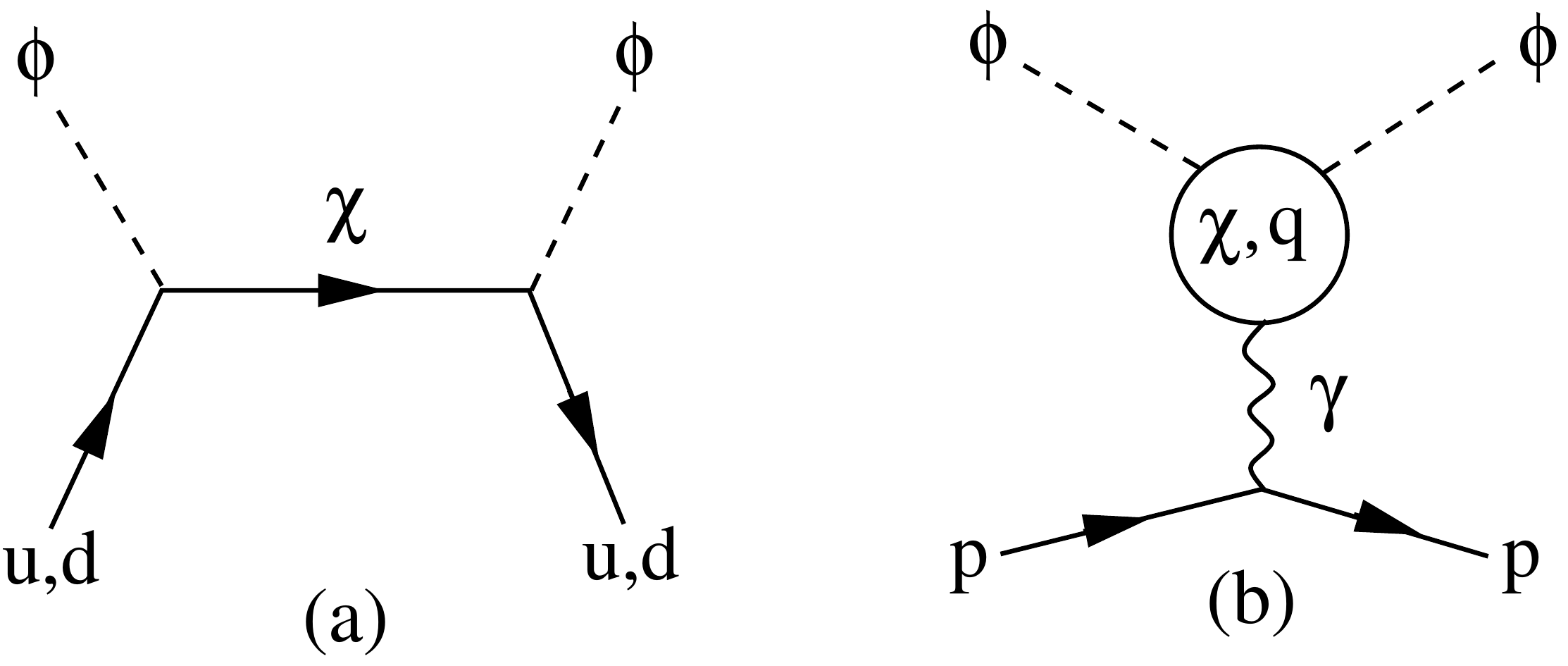}
\end{center}
\caption{SFDM contributions to $\phi$-nucleon scattering via the
  $\phi\bar\chi q$ interaction: (a) tree-level mediator exchange, and
  (b) penguin diagram for DM-proton interaction.
\label{fig:FD_DD}}
\end{figure}

Our dark matter candidate can scatter elastically with quarks through
mediator exchange, fig.\ \ref{fig:FD_DD}(a), leading to DM-nucleon
scattering that is constrained by direct detection experiments.  This
can occur either by the coupling of $\phi$ to valence quarks, or that
to heavy quarks via the loop diagram fig.\ \ref{fig:FD_DD}(b) that
enables photon exchange.  In addition, the Higgs portal coupling
allows for $\phi N\to\phi N$ scattering by Higgs boson exchange.  For
DM masses $m_\phi\gtrsim 6$ GeV, the strongest current limits come
from the LUX experiment \cite{LUX}.  We will first derive constraints
on the different kinds of interactions assuming that they do not
interfere with each other significantly.  In section
\ref{interference} we will consider the possibility of destructive
interference that could weaken direct detection constraints
sufficiently to allow for the indirect signals we discussed in section
\ref{Indirect_detection}.

\begin{figure*}[t]
\centerline{
\includegraphics[width=0.45\textwidth]{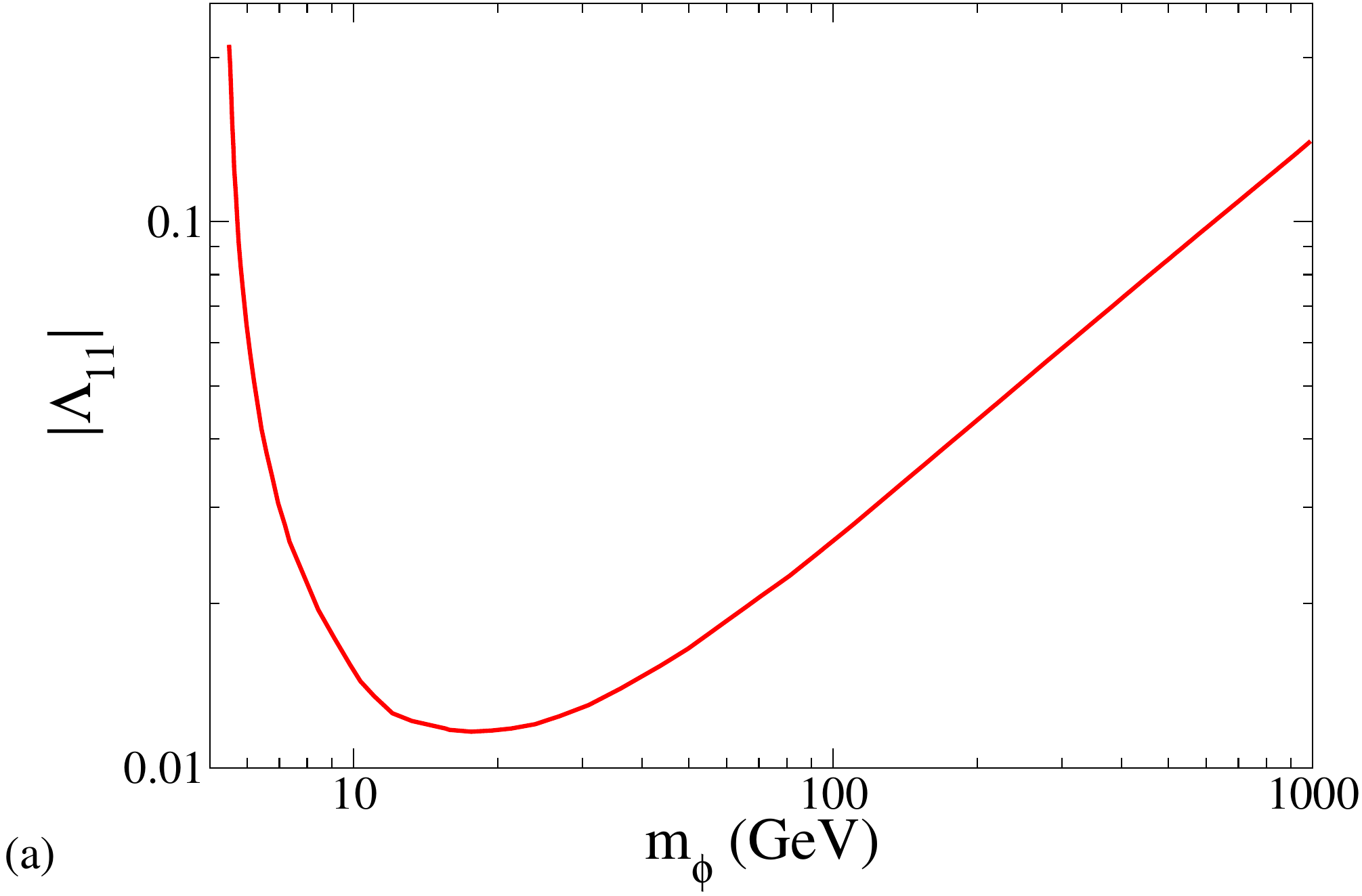}
\includegraphics[width=0.45\textwidth]{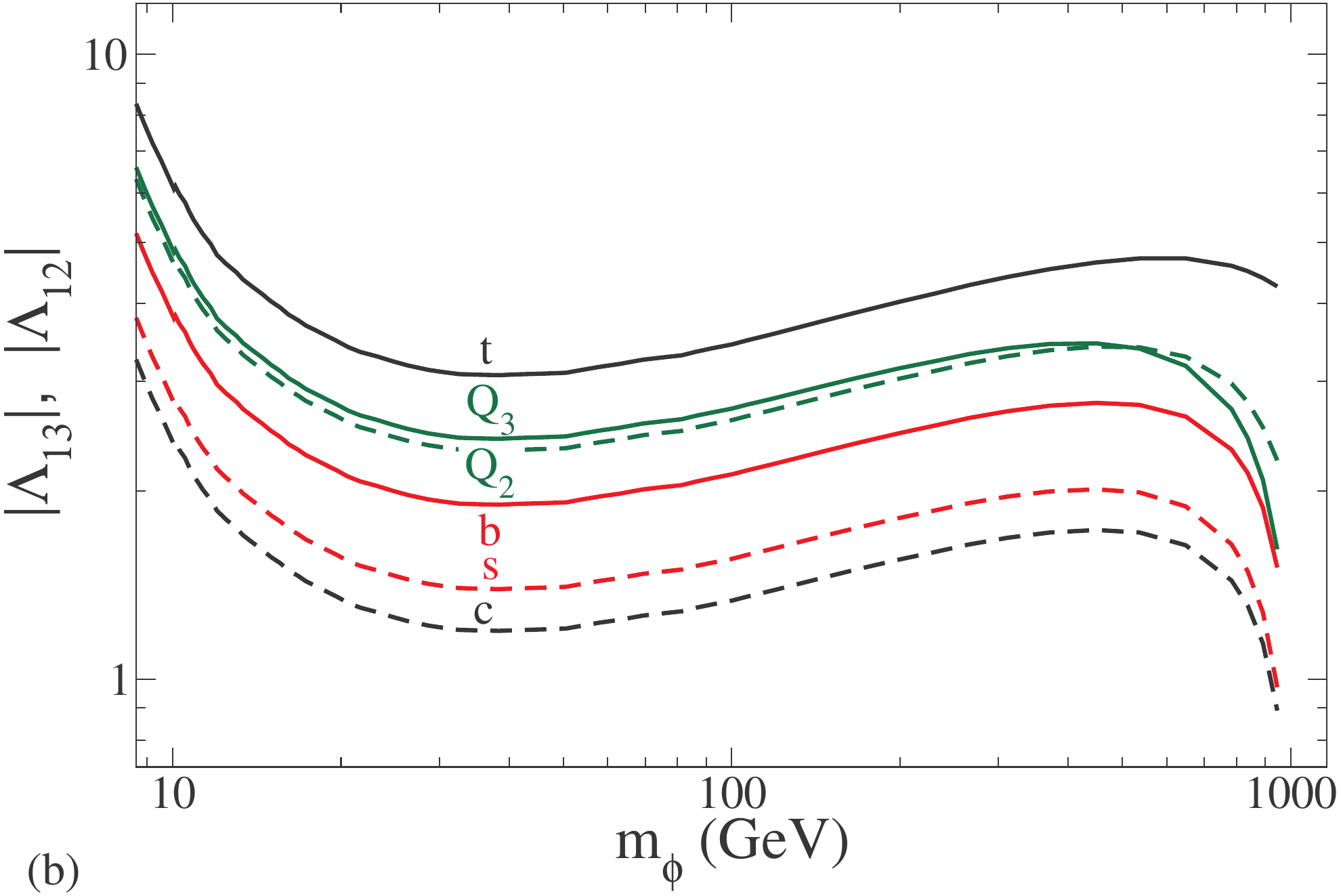}
}
\caption{Left (a): LUX constraints on the coupling of dark matter to
  light quarks $u,d$, assuming mediator mass $m_\chi = 1$ TeV.  Right
  (b): LUX constraints on couplings between 
  $\Lambda_{13}$ (solid curves) and $\Lambda_{12}$ (dashed curves),
  for different choices of the quark appearing in the loop of
  fig.\ \ref{fig:FD_DD}(b), depending upon the choice of model
  ($d$, $u$ or $Q$).  The mediator mass is assumed to be  
	$m_\chi = 1$.}
\label{fig:lux-light-quark}
\end{figure*}

\subsection{Mediator-induced interactions}

We first consider the nonelectromagnetic mediator-induced interaction.
It is straightforward to show that fig.\ \ref{fig:FD_DD}(a) leads to
an effective operator\footnote{Here we ignore contributions suppressed
  by $m_q$ that are irrelevant for $\phi$-nucleon scattering.}
\be
{|\Lambda_{1i}|^2\over
m_\chi^2} (\phi^*\partial_\mu\phi)\, (\bar q_i\gamma^\mu P_{L,R}\,q_i) ~,
\label{effint}
\ee
where the sum over doublet components is taken in the $Q$ model.  When
taking matrix elements of this operator between nucleon states, the
only nonvanishing contributions are from the valence quarks $i=u,d$,
giving the cross section \cite{Fitzpatrick:2010em}
\be
	\sigma_{p,n} = {\mu^2\,f_{p,n}^2\over 4\pi m_\phi^2}
\label{meddd}
\ee
for scattering of $\phi$ on protons or neutrons, where $\mu$ is the
$\phi$-nucleon reduced mass and $f_{p,n}={|\Lambda_{1i}|^2/m_\chi}$,
up to isospin-related factors of order unity, depending upon which DM
model we are considering.\footnote{For $f_p$ these factors are
  $(1,\,1/2,\,3/2)$ for the $u,d,Q$ models, while for $f_n$ they are
  $(1/2,\,1,\,3/2)$, respectively.}  The LUX constraint on these
couplings is shown in fig.\ \ref{fig:lux-light-quark}(a) for mediator
mass $m_\chi = 1$ TeV.  The limit on $\Lambda_{13}$ is orders of
magnitude smaller than values of interest for the relic density for
the coupling to top quarks.  There must be a large generational
hierarchy in the couplings $\Lambda_{1i}$, at least between the first
and higher generations.

Next we consider the contribution from the penguin diagram,
fig.\ \ref{fig:FD_DD}(b).  The loop leads to the effective photon-DM
interaction
\be
	{\kappa\, e\over m_\chi^2 - m_\phi^2} \,(\phi^* {\overset{\leftrightarrow}{\partial}_{\!\mu}}
	\phi)\, 	\partial_\nu F^{\mu\nu} ~.
\ee
For the three models ($u,d,Q$), $\kappa$ is approximately given by
\be
	\kappa \cong {1\over 16\pi^2}\left\{\begin{array}{ll}{Q_{q}} \sum_i
	{|\Lambda_{1i}|^2}\ln
	\left(\frac{m^2_{q_i}}{m^2_\chi}\right),& {\hbox{$q = u$ or $d$}}\\
	\sum_{q,i} Q_{q} {|\Lambda_{1i}|^2}\ln
	\left(\frac{m^2_{q_i}}{m^2_\chi}\right), & \hbox{$Q$ model}
	\end{array}\right.
\ee
in the limit $m_\phi \ll m_\chi$, where $Q_q = 2/3$ or $-1/3$ is the
charge of the quark and $m_{q_i}$ is its mass.  For larger $m_\phi$,
the loop integral depends differently upon $m_\phi$, and the logarithm
gets replaced by
\be
	{1\over(m_\chi^2 - m_\phi^2)}\ln
	\left(\frac{m^2_{q_i}}{m^2_\chi}\right) \to {1\over m_\chi^2} I(\epsilon,\epsilon_i) ~,
\ee
where we define $\epsilon_\phi = m_\phi^2/m_\chi^2$, $\epsilon_i =
m_{q_i}^2/m_\chi^2$, $D(x) = x + \epsilon_i(1-x) - \epsilon_\phi
x(1-x)$, $D'(x) = D(1-x)$ and
\bea
	I &=& \int_0^1 dx\, \Bigg[ (1-x)^3 (1+2x)\left({1\over D'}- {1\over D}\right)
	\nonumber\\
	&+& {(1-x)^4}\left({\epsilon_i + \epsilon_\phi x^2\over 2D^2} - {1 + \epsilon_\phi x^2\over 2 D'^2}
	\right)\Bigg] ~.
\eea
The large logarithm comes from $1/D$ as $x\to 0$.

The resulting photon-mediated DM-proton scattering cross
section is given by
\bea
\sigma_p &=& \frac{16\pi\,\mu^2\alpha^2\kappa^2}{(m_\chi^2-m_\phi^2)^2}~,~~
\eea
where $\mu$ is the $\phi$-proton reduced mass.  The limits from this
process are much weaker than those from the nonelectromagnetic
coupling. Also, whereas that one bounded only $\Lambda_{11}$, this one
applies to $\Lambda_{1i}$ for all the quark generations.  Hence we
take $i=2,3$, (recall that $i=3$ represents the couplings relevant for
the relic density in section \ref{med_dom}).  Ignoring possible
interference between different generations, we obtain the limits shown
in fig.\ \ref{fig:lux-light-quark}(b), with solid (dashed) curves
corresponding to $i=3(2)$.  These couplings are somewhat weaker than
the values leading to the right relic density in
fig.\ \ref{fig:lambda-relic}(a).

\subsection{Higgs portal interaction}

\begin{figure}[t]
\begin{center}
\includegraphics[width=0.45\textwidth]{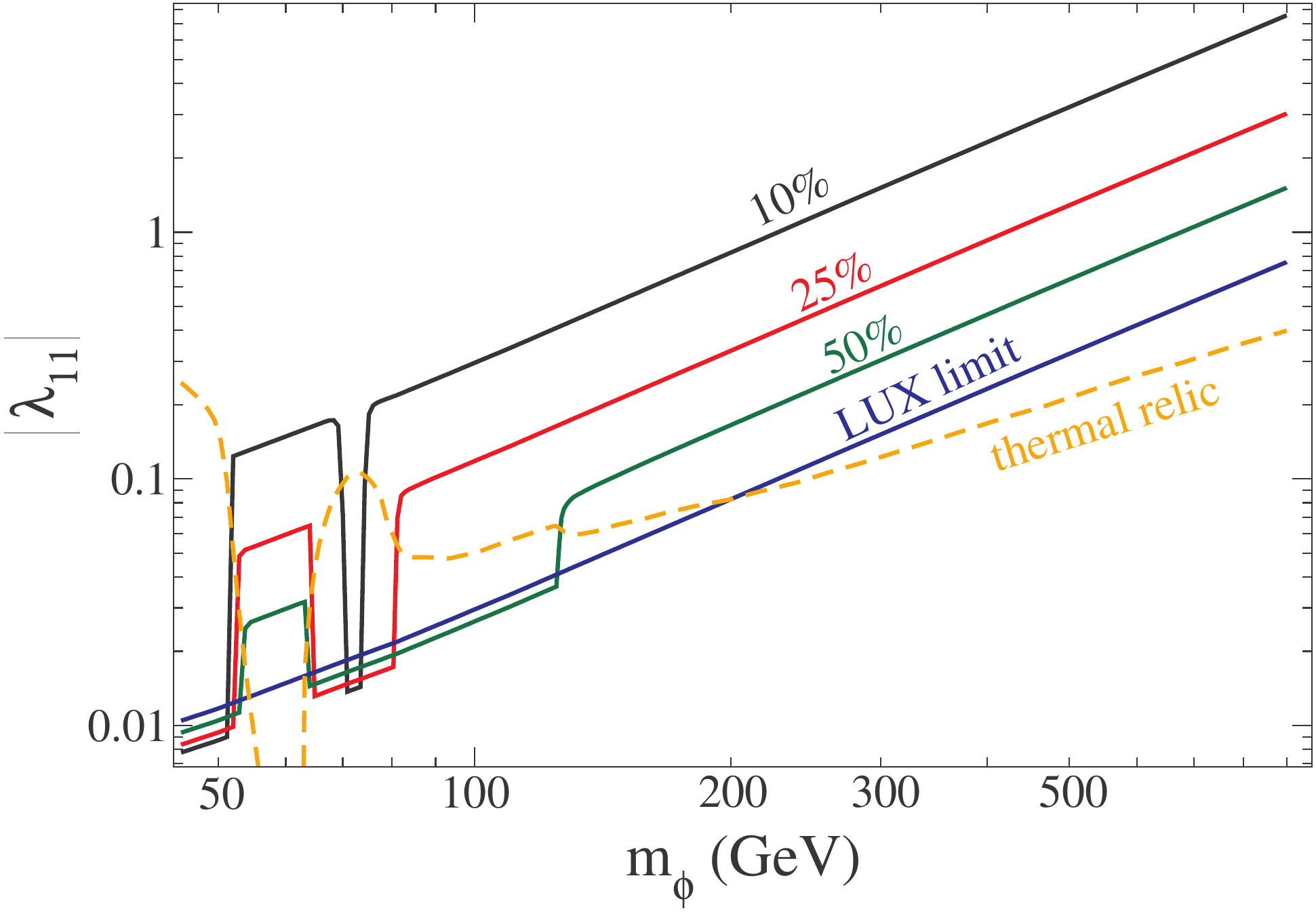}
\end{center}
\caption{Solid curves: LUX and relic density constraints on the Higgs
  portal coupling $\lambda_{11}$, taking dark matter to be
  asymmetric and accounting for accidental cancellation 
  at the level of 100, 50, 25 and 10\% by the
  mediator contribution to the scattering amplitude.
  100\% means no cancellation, giving the usual exclusion curve.
  Dashed curve: value of $\lambda_{11}$ that would give thermal
  relic abundance.
\label{fig:Limits2}}
\end{figure}

For the Higgs portal coupling, the effective DM-nucleon scattering
cross section is given in ref.~\cite{Cline:2013gha}:
\bea
\sigma &=& \frac{\lambda^2_{11}f^2_N}{4\pi}\frac{\mu^2m^2_p}
{m^4_Hm^2_\phi}~,~~
\label{hpdd}
\eea
where $f_N = 0.303$ \cite{Cline:2013gha} is related to the
Higgs-nucleon coupling, and $m_H = 125$ GeV. The LUX data can be used
to put limits on the the coupling $\lambda_{11}$, as shown in
Fig.~\ref{fig:Limits2}.  The value of $\lambda_{11}$ needed for the
observed relic density is also plotted, for the case where only
$\phi_1$ is relevant during freeze-out.  The LUX limit must lie below
the relic density curve if the DM has a thermal origin.  In addition
to the allowed range $m_\phi \gtrsim 150$ GeV, there is a narrow
window of lower masses for which the relic density is not exceeded
around $m_\phi\cong m_h/2$, corresponding to resonantly-enhanced
annihilations.

% Figure 3

\subsection{Higgs-mediator interference}
\label{interference}

Fig.\ \ref{fig:Limits2} shows that the interesting DM mass range for
indirect detection (fig.\ \ref{fig:indirect}) is ostensibly ruled out
by the LUX limit.  However, we have not yet taken into account the
possibility of destructive interference between different
contributions to the $\phi$-nucleon scattering amplitude.  This is
clearly possible for either dark matter particles or their
antiparticles, 
since the effective
operator (\ref{effint})
changes sign under charge conjugation of $\phi$ while the amplitude
from Higgs exchange does not.
Since eqs.\ (\ref{meddd}) and (\ref{hpdd})
have the same dependence upon $m_\phi$, it is particularly simple to
combine them taking account of interference:
\be
	\sigma_n \cong \left(\lambda_{11} f_N {m_p\over m_h^2} \pm 
	{|\Lambda_{11}|^2\over m_\chi} \right)^2 \, {\mu^2\over 4\pi
	m_\phi^2} ~.
\ee 

If interference is destructive for $\phi$ it will be constructive for
$\phi^*$.  Therefore to have a net reduction, it is necessary to
consider asymmetric dark matter where the antiparticle abundance is
suppressed \cite{asymDM}.  The suppression factor has been computed as
a function $r(\lambda/\lambda_0) = n_{\phi^*}/n_\phi$ (the ratio of
anti-DM to DM) in ref.\ \cite{Graesser:2011wi}, where $\lambda_0$
denotes the value of the coupling that would give rise to the correct
thermal relic abundance.\footnote{Denoting $x =
  (\lambda/\lambda_0)^2$, we are able to fit the numerical result of
  ref.\ \cite{Graesser:2011wi} to the function $-\log_{10}r = (A_0 x+
  A_1)/(1 + A_2 x^{A_3})$, where $A_0 = 0.8327$, $A_1=-0.8258$, $A_2 =
  -0.8737$, $A_3 = -0.8213$, which is valid for $x\ge 1$.  For $x<1$,
  $r=1$.}  If the amplitude for scattering of DM on nucleons is
reduced by the factor $(1-\epsilon)$, and that for anti-DM is
increased by $(1+\epsilon)$, and the nominal bound is $\lambda_{\rm
  LUX}$, then the relaxed bound on the coupling is given by
\be
	\lambda_{\rm eff} = {\lambda_{\rm LUX}\over
	\left[(1-\epsilon)^2(1-r/2) +
        (r/2)(1+\epsilon)^2\right]^{1/2}}
\label{leff}
\ee
where $r = r(\lambda_{\rm eff}/\lambda_0)$.  Eq.\ (\ref{leff}) gives
only an implicit definition of $\lambda_{\rm eff}$, but it can be
solved numerically by iteration.

In fig.\ \ref{fig:Limits2} we show the modified upper limits 
on $|\lambda_{11}|$ that result from allowing for 
accidental cancellations that reduce the amplitude to $75\%$, $50\%$
and $10\%$ of its magnitude in the absence of the mediator
contribution.  It is clear that the range of allowed masses can 
be considerably widened relative to the thermal abundance 
scenario. 

The mediator diagram can have a significant effect only for $\phi$-nucleon
scattering, and not $\phi\phi^*$ annihilation, because of the quark
vector current in the effective interaction (\ref{effint}).  Its
matrix element for $\phi\phi^* \to q\bar q$ is suppressed by $m_q =
m_b$ for the mass range of interest, while that for $\phi N\to \phi N$
suffers from no such kinematic reduction.

\section{DM-induced flavor effects}
\label{FCNC_effects}

We now turn to the implications of scalar flavored dark matter for
particle-physics phenomenology, including FCNC processes, rare decays,
and CP violation.  We recall our choice of the underlying flavor
symmetry group as
SU(3)$_\phi\times$SU(3)$_Q\times$SU(3)$_u\times$SU(3)$_d$
\cite{Agrawal:2014aoa}, which is broken by the SM Yukawa couplings and
our new couplings $\Lambda_{ij}$.  Because the mediator $\chi$ is
forced to be heavy by LHC constraints, we do not need to rely upon a
more restrictive flavor structure such as MFV \cite{D'Ambrosio:2002ex}
to keep flavor-changing neutral currents under control, as will become
clear in this section.  However, we will demonstrate the potential of
the model to give rise to observable low-energy effects for values of
the couplings $\Lambda_{ij}$ that are consistent with the constraints
obtained in the previous sections.

\begin{figure*}[t]
\centerline{
\includegraphics[width=0.7\textwidth]{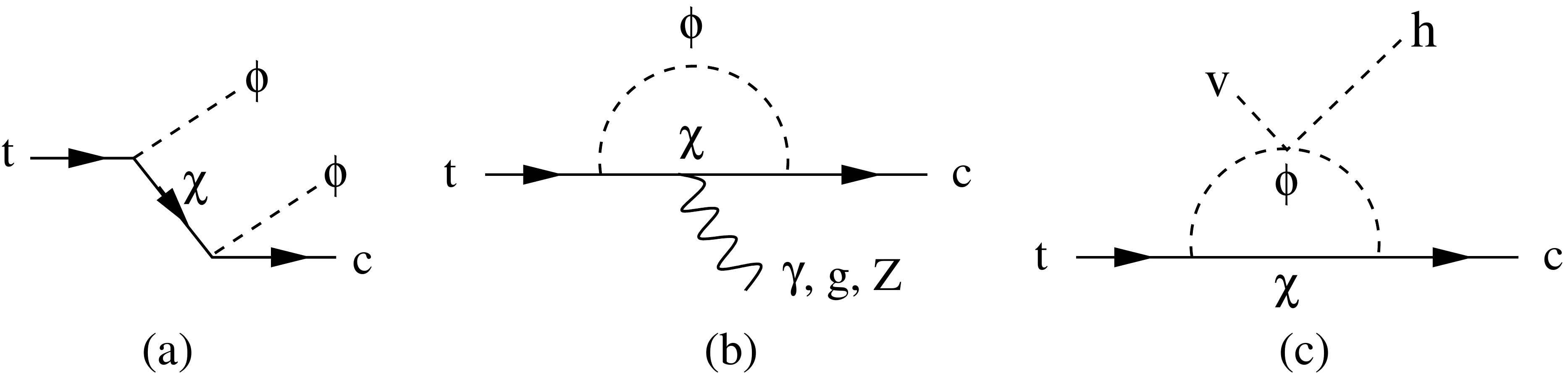}
\includegraphics[width=0.2\textwidth]{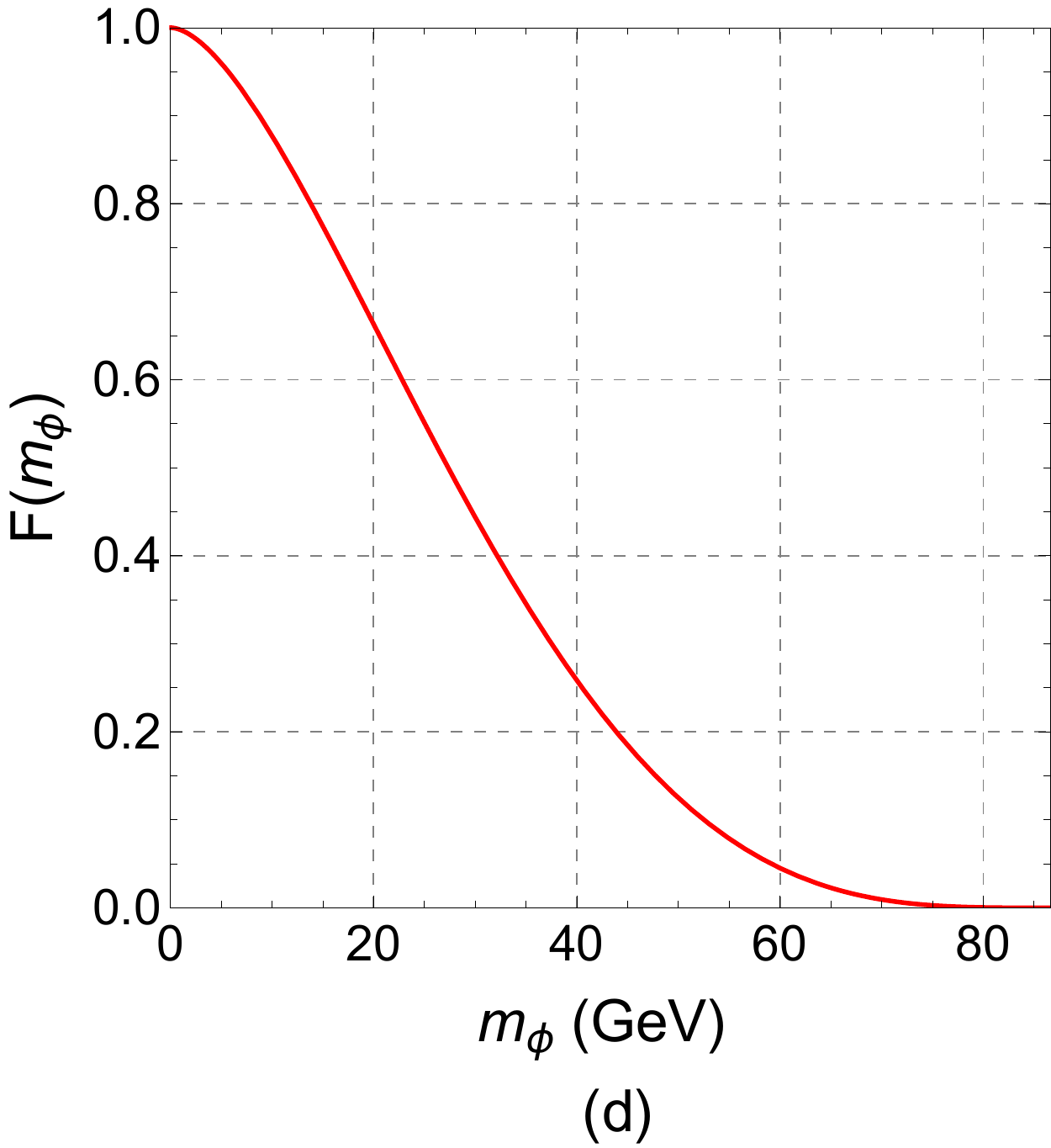}}
\caption{Diagrams for (a) $t \to \phi\phi^* c$, (b) $t\to c\gamma$ or
  other gauge bosons, (c) $t\to h c$, and (d) the function $F(m_\phi)$
  determining the partial width for $t \to \phi\phi^* c$ relative to
  its maximum value.}
\label{fig:decays}
\end{figure*}

\subsection{Flavor-changing meson oscillations}
\label{Bmixing}

We briefly review the formalism for $\Delta F=2$ flavor-changing
oscillations of neutral mesons.  To be concrete, we illustrate this
for the case of $\Delta B = 2$ meson mixing.  In the $B^0$-$\bar B^0$
basis, the mixing is described by the $2 \times 2$ matrix $M -
\frac{i}{2} \Gamma$, in which the mass ($M$) and decay ($\Gamma$)
matrices are Hermitian. The physical states are \cite{DR}
\bea
\ket{B_L} &=& p \ket{B^0} + q \ket{\bar B^0} ~, \nn\\
\ket{B_H} &=& p \ket{B^0} - q \ket{\bar B^0} ~,
\eea
with eigenvalues ($L=$ ``light,'' $H=$ ``heavy'')
\be
\mu_{L,H} = M_{L,H} - \frac{i}{2} \Gamma_{L,H} ~,
\ee
in which $M_{L,H}$ and $\Gamma_{L,H}$ denote the masses and decay
widths of $B_{L,H}$. In addition,
\be
\frac{q}{p} = \pm \left( \frac{M_{12}^* - i \Gamma_{12}^* / 2}{M_{12} - i \Gamma_{12} / 2} \right)^{1/2} ~.
\label{qoverp}
\ee
It is a good approximation to take $|\Gamma_{12}| \ll |M_{12}|$;
then
\be
\Delta M \equiv M_H - M_L \cong 2 \left\vert M_{12} \right\vert .
\ee
In our model, the matrix element $M_{12}$ receives new contributions
beyond the SM from box diagrams with $\phi$ and $\chi$ in the loop.

\subsubsection{$B_s^0$-$\bar B_s^0$ mixing}

The DM-induced contributions to $B_s^0$-$\bar B_s^0$ mixing from box
diagrams with $\phi$ and $\chi$ in the loop can be described by the
effective operator \cite{Agrawal:2014aoa}
\be
	{(\Lambda^\dagger\Lambda)_{bs}^2\over 128\pi^2\, m_\chi^2}
	\left(\bar b\,\gamma^\mu P_{L,R}\, s\right)^2
\label{DMBsmixing}
\ee
in the $Q$- ($P_L$) and $d$-type ($P_R$) models, where we used the
approximation $m_{\phi_i}\ll m_\chi$.\footnote{This follows from
  eq.\ (4.1) of ref.\ \cite{Agrawal:2014aoa}, accounting for the loop
  now being dominated by momenta of order $m_\chi$, and ignoring
  corrections of order $(m_{\phi}/m_\chi)^2$.}
The corresponding mass splitting is $(\Delta M_s)_{DM} =
|(\Lambda^\dagger\Lambda)_{bs}^2| m_{B_s} f_{B_s}^2 /(192\pi^2\,
m_\chi^2)$. The measured value is $(\Delta M_s)_{\rm exp.} = (11.69\pm
0.02)\times 10^{-9}$ MeV, while the SM prediction is $(\Delta
M_s)_{\rm SM} = (11.4 \pm 1.7)\times 10^{-9}$ MeV \cite{Lenz}. These
quantities are related via
\be
(\Delta M_s)_{\rm exp.} = 2 \left\vert (M_{12})_{\rm SM} + (M_{12})_{\rm DM} \right\vert ~.
\label{DMs}
\ee

To obtain constraints on the DM contribution to (\ref{DMs}), one has
to take into account a possible phase difference between
$(M_{12})_{\rm SM}$ and $(M_{12})_{\rm DM}$. But this phase difference
will also manifest itself in $q/p$, eq.\ (\ref{qoverp}). A rigorous
analysis would require a simultaneous fit to the measured values of
$\Delta M_s$ and arg$(q/p)$, which is beyond the scope of this
paper. Instead, to estimate the allowed size of the new contribution,
we neglect any phase difference, leading to $|\Delta M_s|_{\rm DM} =
(0.3 \pm 1.7)\times 10^{-9}$ MeV, or $|\Delta M_s|_{\rm DM} \le 5.4
\times 10^{-9}$ MeV at 3$\sigma$. For $m_\chi = 1$ TeV, this
corresponds to the limit $|(\Lambda^\dagger\Lambda)_{bs}|< 0.19$.
This is smaller than the direct detection bounds on second-generation
couplings shown in fig.\ \ref{fig:lux-light-quark}(b).

As can be seen from the above values of $(\Delta M_s)_{\rm exp.}$ and
$(\Delta M_s)_{\rm SM}$, the measurement of $B_s$-$\bar B_s$ mixing is
consistent with the SM prediction. On the other hand, the theoretical
error on this prediction is sizeable, leaving ample room for a
new-physics contribution to $\Delta M_s$. Indeed, if $|\Delta
M_s|_{\rm DM}$ saturates its upper limit, it will be of the same order
as $|(\Delta M_s)_{\rm SM}|$.  We therefore see that flavored DM could
contribute significantly to $B_s$-$\bar B_s$ mixing with reasonable
values of the couplings.

\subsubsection{$K^0$-$\bar K^0$, $D^0$-$\bar D^0$, $B_d^0$-$\bar B_d^0$ mixing}

A similar analysis can be done for oscillations of the other neutral
meson systems, $K^0$-$\bar K^0$, $D^0$-$\bar D^0$, $B_d^0$-$\bar
B_d^0$.  Constraints on the coefficients $c_{ij}$ of the effective
operator $\Lambda^{-2}(\bar q_i \gamma_\mu P_{\sss L} q_i)^2$ (where
$\Lambda$ is the new physics scale) have been compiled for example in
ref.\ \cite{Isidori:2010kg}. These can be related to the prediction
(\ref{DMBsmixing}), with appropriate substitution of quark flavors.
The results are shown in table \ref{tab:mixing}. For $K^0$-$\bar K^0$
and $D^0$-$\bar D^0$ mixing, we obtain separate constraints on the
real and imaginary parts of $(\Lambda\Lambda^\dagger)^2_{ij}$ ($ij =
ds, uc$). For $B_d^0$-$\bar B_d^0$ and $B_s^0$-$\bar B_s^0$ mixing,
the constraints are given only for $|(\Lambda\Lambda^\dagger)^2_{ij}|$
($ij = bd, bs$). The imaginary parts of
$(\Lambda\Lambda^\dagger)^2_{ij}$ can lead to CP-violating effects, as
we will discuss in section \ref{cpviolation}.

\begin{table}[t]
\begin{tabular}{|c|c|c|}
\hline
$ij$ & Re$\left[ (\Lambda\Lambda^\dagger)^2_{ij} \right]$ & Im$\left[ (\Lambda\Lambda^\dagger)^2_{ij} \right]$ \\
\hline
$ds$ & $1.1\times 10^{-3}$ & $4.3\times 10^{-6}$ \\
\hline
$uc$ & $7.1\times 10^{-4}$ & $1.3\times 10^{-4}$ \\
\hline
$bd$ & \multicolumn{2}{|c|}{$3.6 \times 10^{-4}$} \\
\hline
$bs$ & \multicolumn{2}{|c|}{$8.3 \times 10^{-3}$} \\
\hline
\end{tabular}
\caption{Bounds on FCNC matrix elements of
  $(\Lambda\Lambda^\dagger)_{ij}$ with $i\neq j$ from neutral meson
  mixing, assuming mediator mass $m_\chi = 1\,$TeV. Values for the first
  two rows are inferred using constraints reported in ref.\ 
  \cite{Isidori:2010kg}.  For $B_{d,s}$ (last two rows) we constrain 
  only the modulus $|(\Lambda\Lambda^\dagger)^2_{ij}|$ using updated 
  experimental and SM fit numbers from \cite{lenzpc}.}
\label{tab:mixing}
\end{table}

\subsection{Flavor-changing top quark decays}

SFDM allows for a variety of rare FCNC decays, including $t\to c(u)
\phi\phi$ (if $m_\phi$ is sufficiently small), $b\to s \gamma$, $t \to
(Z,h,g,\gamma) c$, and $(h,Z) \to b\bar s$.  A summary of the Fermilab
and LHC constraints on these processes is given in
ref.\ \cite{Goldouzian:2014xfa}.  With the exception of $t\to c
\phi\phi$, these are unobservably small, despite having no symmetry
(MFV) to suppress them.  This is a consequence of the chiral structure
of the interaction (\ref{couplings}), which causes all amplitudes to
be suppressed by $1/m_\chi^2$ and not just $1/m_\chi$.

If $m_\phi \lesssim m_t/2$, the tree-level processes $t\to c \phi\phi$
or $t\to u \phi\phi$ are allowed (fig.\ \ref{fig:decays}(a)).  For the
$c\phi\phi$ final state, the partial width is
\bea
	\delta\Gamma &\cong& {(\Lambda^\dagger\Lambda)_{tt} 
(\Lambda^\dagger\Lambda)_{cc}\, m_t^5\over 4096\,\pi^3\, m_\chi^4}
	F(m_\phi) \nonumber\\
	&=& 2\times 10^{-6}\,{\rm GeV}\cdot|\Lambda|^2_{tt}
|\Lambda|^2_{cc}\,
F(m_\phi) ~,
\label{tcphiphi}
\eea
where the dependence on $m_\phi$ is shown in fig.\ \ref{fig:decays}(d)
and for the numerical estimate we took $m_\chi = 900\,$ GeV.  The
analogous formula with $c\to u$ applies for $t\to u \phi\phi$, but
because of the more stringent constraint on first-generation couplings
from direct detection, this is expected to be subdominant.  With large
couplings $\Lambda\sim 3$ and light DM with $m_\phi\sim 30\,$GeV,
eq.\ (\ref{tcphiphi}) would lead to a branching ratio of $3\times
10^{-5}$.  Recent studies of this process in other models with
flavor-changing scalar DM coupling to the top estimate that LHC
searches could ultimately be sensitive to such a small branching ratio
\cite{Li:2011ja,He:2007tt,Jia:2015uea}.  Although our choice of
$m_\phi$ is ruled out by direct detection for a thermally produced
WIMP, since $\phi$ has a conserved particle number, there could be a
DM asymmetry allowing for sufficiently small coupling to the Higgs for
consistency with direct searches.\footnote{This would also require
  some fine tuning of the loop contributions to $\lambda_{ij}$,
  according to the considerations of section \ref{when}.}

A second class of decays is $t\to c$ $+$ gauge boson, shown in
fig.\ \ref{fig:decays}(b).  ATLAS obtains upper limits of $1.7 \times
10^{-4}$ on the branching ratio for $t \to c g$ and $4\times
10^{-5}$ for $t\to u g$ \cite{Aad:2012gd,Aad:2015gea}.  Writing the NP
contribution to the $t \to c$ chromomagnetic dipole moment as
\be
g_s \, {\kappa_{tcg}} \, 
(\bar t\, \sigma^{\mu\nu} T^a P_{L,R}\, c)\,  G^a_{\mu\nu} + {\rm
h.c.}~,
\ee
the limit on the branching ratio corresponds to $\kappa_{tcg}
< 1.3 \times 10^{-2}$ TeV$^{-1}$. In our model
\be
{\kappa_{tcg}} = (\Lambda^\dagger\Lambda)_{tc} \, \frac{m_t}{64\pi^2 m_\chi^2} ~,
\ee
implying the weak constraint $(\Lambda^\dagger\Lambda)_{tc} <
40$.

For the electromagnetic FCNC $t\to u\gamma$ decays, CMS finds a limit
of $1.6\times 10^{-4}$ $(1.8\times 10^{-3}$ for $t\to c\gamma$)
\cite{CMS:2014hwa}.  This corresponds to a limit on the magnetic
moment coefficient $\kappa_{tu\gamma}$
\be
{2e\over 3} \, {\kappa_{t u \gamma}} \, 
(\bar t\, \sigma^{\mu\nu}  P_{L,R}\, u)\,  F_{\mu\nu} + {\rm
h.c.}
\ee

of $\kappa_{tu\gamma} < 0.16\,$TeV$^{-1}$, and a correspondingly
weaker limit of $(\Lambda^\dagger\Lambda)_{tu}< 580$.  The best limit
on $t\to qZ$ also comes from CMS, with an upper bound of BR $< 5\times
10^{-4}$ \cite{Chatrchyan:2013nwa}, leading to
$(\Lambda^\dagger\Lambda)_{tc} + (\Lambda^\dagger\Lambda)_{tu}\lesssim
785$ for models with $Q$-like mediators, and somewhat less stringent
for $u$-like.

The decay mode $t\to c h$ shown in fig.\ \ref{fig:decays}(c) has
a partial width of order
\bea
	\delta\Gamma &\cong& 
	\left(v(\Lambda^\dagger\lambda\Lambda)_{tc}\over
	16\pi^2\, m_\chi^2\right)^2 {m_t (m_t^2-m_h^2)\over 16\pi}
	\nonumber\\
	&\cong& 2\times 10^{-7} (\Lambda^\dagger\lambda\Lambda)_{tc}^2
	{\rm\, GeV} ~,
\eea
which is far below the current sensitivity of $\delta\Gamma \lesssim
1{\rm\, GeV}$ \cite{Khachatryan:2014jya} for reasonable value of the
couplings.

\subsection{Flavor-changing $b$ decays}

The radiative flavor-changing processes $b\to s\gamma$  and $b\to
s \ell^+\ell^-$ are described by the effective operators
\bea
{\cal O}_7 &=& \frac{e}{(4\pi)^2} m_b ({\bar s} \sigma^{\mu\nu} P_R b) F_{\mu\nu} ~, \nn\\
{\cal O}_9 &=& \frac{e^2}{(4\pi)^2} ({\bar s} \gamma_\mu P_L b) ({\bar\ell} \gamma^\mu \ell) ~, \nn\\
{\cal O}_{10} &=& \frac{e^2}{(4\pi)^2}({\bar s} \gamma_\mu P_L b) 
({\bar\ell} \gamma^\mu \gamma_5 \ell)
\label{ops}
\eea
and their chirality-flipped counterparts,  ${\cal O}'_7$, ${\cal O}'_9$ and ${\cal
  O}'_{10}$,
obtained by taking $P_L \leftrightarrow P_R$. The operators ${\cal O}_{7,9,10}$ are induced by
the $Q$ model, while the $d$ model generates ${\cal O}'_{7,9,10}$.

\subsubsection{$b\to s\gamma$}

The decay $b\to s\gamma$ has reduced sensitivity because of both loop
and chiral suppression of the induced magnetic moment operator. In the
$Q$ model, to leading order in $1/m^2_\chi$, it is given by the
diagram analogous to fig.\ \ref{fig:decays}(b),
\be
	 {Q_b (\Lambda^\dagger\Lambda)_{bs} \over
	12\, m_\chi^2}\,{\cal O}_7 \equiv {4 G_F\over\sqrt{2}}
V_{tb} V^*_{ts}\, C^{\rm\sss DM}_7\, {\cal O}_7
\ee
where $Q_b = -1/3$ is the charge of the $b$ quark, and 
$C^{\rm\sss DM}_7$ denotes the new DM contribution.

In ref.\ \cite{Descotes-Genon:2013wba}, a global analysis of $B$ decay
processes was performed, motivated by tensions with the SM predictions
revealed by LHCb measurements \cite{Aaij:2013qta}.  There it was shown
that a nonzero contribution from new physics is preferred at 1$\sigma$
for ${\cal O}_7$, namely $C^{\rm\sss DM}_7 \in [-0.05,-0.01]$.  For
$m_\chi = 1$ TeV, $0.5 < (\Lambda^\dagger\Lambda)_{bs} < 2.4$ is the
corresponding allowed range of couplings.

A similar analysis can be performed for the Wilson coefficient $C'_7$
which appears in the $d$ model, where the 1$\sigma$ range is
$C^{'\rm\sss DM}_7 \in [-0.04,0.02]$ This corresponds to the range of
couplings $-0.9<(\Lambda^\dagger\Lambda)_{bs}<1.9$.  From both $C^{\rm
  DM}_7$ and $C^{'\rm DM}_7$ the constraints on
$(\Lambda^\dagger\Lambda)_{bs}$ are much weaker than the limit
$|(\Lambda^\dagger\Lambda)_{bs}|< 0.19$ we obtained above from
$B_s$-$\bar B_s$ mixing. Hence if one imposes the $B_s$-$\bar B_s$
mixing constraint, our DM model cannot generate large enough
contributions to $B\to X_s\gamma$ decays to address the current (weak)
hints of deviations from the SM predictions.

\subsubsection{$b\to s\ell^+\ell^-$}

The new DM interactions contribute to $b \to s \ell^+ \ell^-$ through
$b \to s \gamma^* (\to \ell^+ \ell^-)$ or $b \to s Z^* (\to \ell^+
\ell^-)$ (i.e., the $\gamma$ or $Z$ is off-shell).  All three of the
operators in (\ref{ops}) (or their chirally-flipped counterparts) can
be relevant.

There has been a great deal of activity, both theoretical and
experimental, concerning $B \to K^{(*)} \mu^+ \mu^-$ decays; see
ref.\ \cite{Matiasetal} for a recent review.  At present, there is a
hint of new physics in $C_9$: at $1\sigma$, it is found that $C_9^{\rm
  NP} \in [-1.6,-0.9]$, and remains nonzero even at $3\sigma$
\cite{Descotes-Genon:2013wba}.  Within the $Q$ ($d$) model, we find
that the $b \to s \gamma^*$ contribution to $C_9$ $(C'_9)$ is
\be
C_9^{\rm\sss DM}\  (C_9^{'\rm\sss DM}) = \frac{7\sqrt{2} Q_b(\Lambda^\dagger\Lambda)_{bs}}{144\,
 G_F \,
m^2_\chi\, |V_{tb}||V_{ts}|} ~.
\ee

at leading order in $1/m_\chi^2$.  For $m_\chi = 1$ TeV, the range of
couplings $(\Lambda^\dagger\Lambda)_{bs} \in[18,\,33]$ corresponds to
the 1$\sigma$ range of $C_9^{\rm\sss NP}$. Similarly to the case of
$b\to s\gamma$, this is two orders of magnitude larger than the
constraint from $B_s$-$\bar B_s$ mixing; hence the DM contribution
cannot explain the discrepancy in $C_9$.  For $C_9^{'\rm\sss DM}$ the
1$\sigma$ allowed range is $[-0.2,0.8]$, again corresponding to
constraints on $(\Lambda^\dagger\Lambda)_{bs}$ that are much weaker
than those from $B_s$-$\bar B_s$ mixing.

Similar conclusions hold for all $b \to s \ell^+ \ell^-$ and $b \to s
q \bar q$ operators.  The DM contribution to the Wilson coefficients
is suppressed relative to the SM by $O(M_W^2/m_\chi^2) \sim 1\%$. It
cannot be compensated by large values of
$(\Lambda^\dagger\Lambda)_{bs}$, due to the constraint from
$B_s$-$\bar B_s$ mixing.

\begin{figure}[t]
\centerline{
\includegraphics[width=0.3\textwidth]{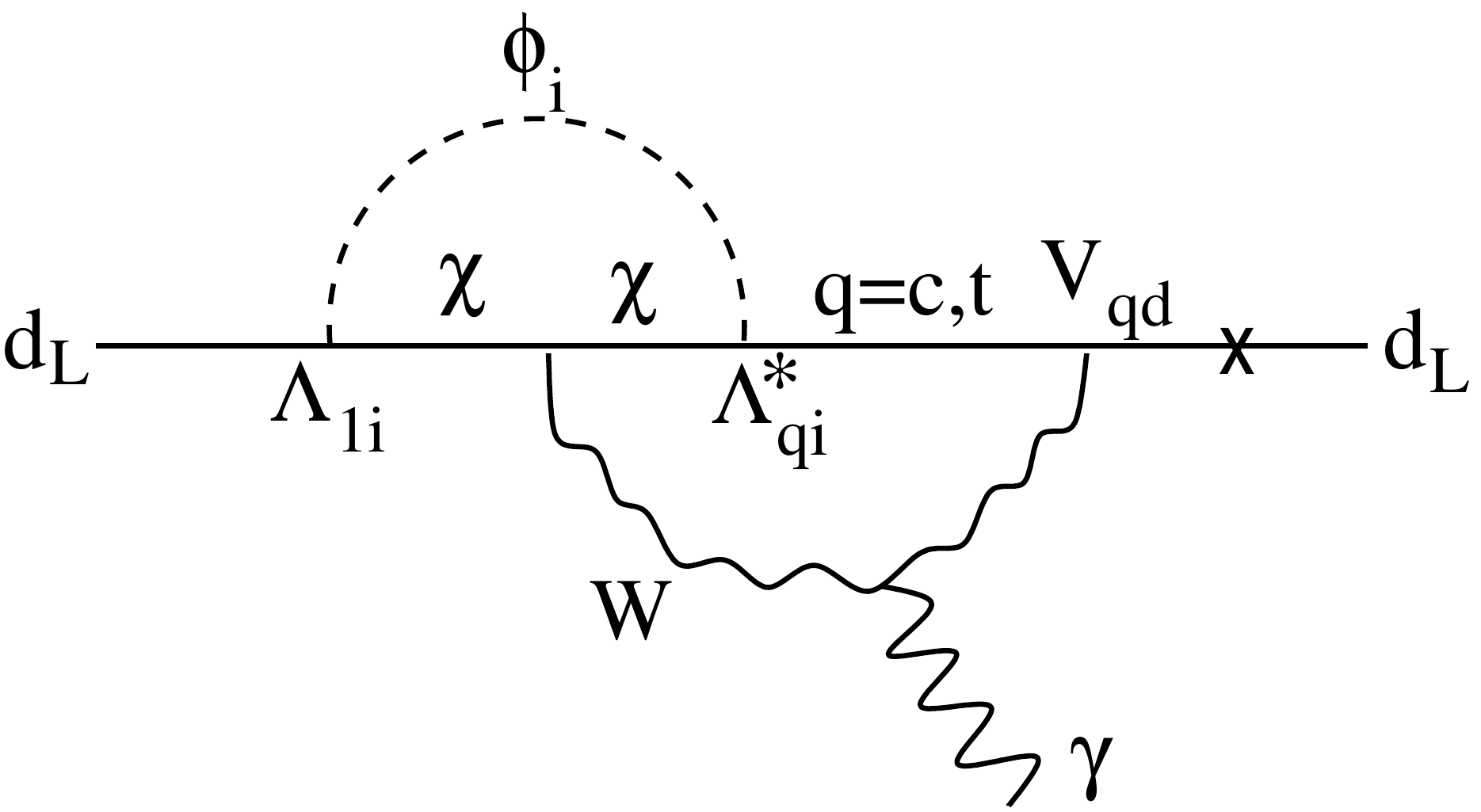}
}
\caption{Diagram giving down-quark electric dipole moment.}
\label{fig:edm}
\end{figure}

\subsection{CP violation}
\label{cpviolation}

The couplings $\Lambda_{ij}$ in our model can be complex, leading to
new sources of CP violation.  They can have observable effects through
meson mixing in $B^0$ decays, and possibly also through the electric
dipole moment of the neutron.

\subsubsection{Indirect CP violation in $B^0$ decays}

In sec.\ \ref{Bmixing} we showed that the DM-induced contribution to
$B_s$-$\bar B_s$ mixing may be significant for reasonable values of
the couplings $(\Lambda^\dagger\Lambda)_{bs}^2\sim 0.1$ in
eq.\ (\ref{DMBsmixing}).  The imaginary part of
$(\Lambda^\dagger\Lambda)_{bs}^2$ is a new source of CP violation,
entering in $q/p$, eq.\ (\ref{qoverp}),
\be
\frac{q}{p} \cong \pm \left(\frac{(M_{12})^*_{\rm SM} + (M_{12})^*_{\rm DM}}{(M_{12})_{\rm SM} + (M_{12})_{\rm DM}}\right)^{1/2} ~.
\ee
which is approximately a pure phase, $|q/p|\cong 1$.  In the SM, $q/p
= V_{tb}^* V_{ts} / V_{tb} V_{ts}^*$.

The phase arg$(q/p)$ can be measured using indirect CP violation in
$B_s$ decays. The main experimental focus has been on $\phi_s^{c{\bar
    c}s}$, which is the phase as measured via indirect CP asymmetries
in $B_s$ decays with $b \to c{\bar c}s$ ($B_s \to J/\psi \phi$,
$J/\psi K^+ K^-$, $J/\psi \pi^+ \pi^-$, $D_s^+ D_s^-$).  Its predicted
value is $\phi_s^{c{\bar c}s} = {\rm arg}(q/p) =
-0.0363^{+0.0012}_{-0.0014}$ in the SM, while the measured value is
$-(0.015 \pm 0.035)$ \cite{HFAG}. Although these are consistent with
one another, there is ample room for a new-physics contribution.
Given that $(M_{12})^*_{\rm DM}$ can be of the same order as
$(M_{12})^*_{\rm SM}$, our model could contribute significantly to
$\phi_s^{c{\bar c}s}$.

\subsubsection{CP-violation in \
$K^0$-$\bar K^0$, $D^0$-$\bar D^0$, $B_d^0$-$\bar B_d^0$ mixing}

In contrast to the $B^0_s$ system, the CP phase relevant for
$K^0$-$\bar K^0$ and $D^0$-$\bar D^0$ mixing is sufficiently
well-measured to provide a separate constraint on the imaginary part
of $(\Lambda^\dagger\Lambda)_{ij}$ for the off-diagonal elements.  The
upper limits are given in table \ref{tab:mixing}. For $B_d^0$-$\bar
B_d^0$ mixing, we do not present a constraint on the imaginary part of
$(\Lambda^\dagger\Lambda)_{ij}$.  However, its modulus is reasonably
well-constrained, so that its imaginary piece cannot be too large. We
therefore do not expect significant DM-induced contributions to CP
violation in $K^0$-$\bar K^0$, $D^0$-$\bar D^0$, or $B_d^0$-$\bar
B_d^0$ mixing.

\subsubsection{Electric dipole moments}

The new phases can also produce quark electric dipole moments through
two-loop graphs like that shown in fig.\ \ref{fig:edm}.  The extra
loop with $W$ exchange is needed to get the products
$(\Lambda\Lambda^\dagger)_{1j}$ with $j\neq 1$, since there are no
phases in $(\Lambda\Lambda^\dagger)_{11}$.  Because of chiral
suppression, the resulting quark EDM is small,
\bea
	d_d &\sim& {{\rm Im}[(\Lambda\Lambda^\dagger)_{12}]\, e\,
	 g_2^2\, V_{cd}\,
	m_d\over (16\pi^2\, m_\chi)^2} \nonumber\\
	&\cong& 3\times 10^{-28}\, {\rm Im}[(\Lambda\Lambda^\dagger)_{12}]
	\,e\cdot{\rm cm}.
\eea
Given the stringent constraints on $(\Lambda\Lambda^\dagger)_{uc}$
and  $(\Lambda\Lambda^\dagger)_{ds}$
from $D^0$-$\bar D^0$ and $K^0$-$\bar K^0$ mixing, this is negligible compared to
the current sensitivity through the neutron EDM, 
$3\times 10^{-26}\,e\cdot $cm.

\begin{table}[t]
\begin{tabular}{|c|c|r|c|c|}
\hline
& $m_\phi$ (GeV) & $\lambda_{11}$ & $\Lambda_{ij}$ & comment \\
\hline 
1& 60 & 0.01 & $-$ & asymmetric dark matter\\
2& 63 & $0.016$ & $-$ & GC excess\\
3& 100 & $-0.12$ & $\Lambda_{11}=0.04$   & direct detection interference \\
4& 200 & 0.08 & $-$ & thermal relic/Higgs portal\\
5& 700 & 0.27 & $|\Lambda_{13}|=0.8$ & thermal relic/mediator\\ 
\hline
\end{tabular}
\caption{Parameter values for benchmark models, assuming
$m_\chi = 1\,$TeV.  Dashes indicate where $\Lambda_{ij}$
can take a range of values.}
\label{tab:bm}
\end{table}

\section{Benchmark models}
\label{benchmarks}

Rather than trying to combine the constraints we have discussed to
obtain allowed regions in parameter space, since we have many
parameters, here we will instead give a few examples of allowed
parameter values that illustrate the different qualitative
possibilities of the model.  For simplicity, we fix the mediator mass
$m_\chi = 1\,$TeV, close to the lower limit from LHC searches.  The
benchmark models are summarized in table \ref{tab:bm}.

Model 1 underscores the fact that if the annihilation cross section
exceeds that needed for the thermal relic density, then we could
appeal to the complex nature of SFDM to assume that it has an
asymmetry accounting for its abundance.  The example chosen here has
$\lambda_{11}$ close to the upper limit from direct detection
searches, making such a model close to discovery.

Model 2 is the largest $m_\phi$ below $200$ GeV allowed by direct
detection and the thermal relic value of $\lambda_{11}$ shown in
fig.\ \ref{fig:Limits2}.  It is marginally consistent with the
galactic center gamma ray excess, fig.\ \ref{fig:indirect}, though in
conflict with the Fermi dwarf spheroidal constraints.

Model 3 illustrates a case that would be ruled out for thermal DM but
is allowed for asymmetric DM due to destructive interference between
Higgs and mediator exchange contributions to DM-nucleon scattering.

Model 4 is an example where the thermal relic density arises from
Higgs portal interactions, and $\lambda_{11}$ is close to the LUX
limit in fig.\ \ref{fig:Limits2}, again making this model detectable
by the next improvement in sensitivity.

Model 5 is chosen to illustrate the window of couplings shown in
fig.\ \ref{fig:lambda-comp} where annihilation of dark matter by
mediator exchange can dominate over the Higgs portal coupling, without
fine tuning of parameters.  At this mass, the LUX limit upper
$\lambda_{11}<0.5$ is satisfied.  Moreover $|\Lambda_{13}|$ is below
the direct detection limit $\sim 2$ shown in
fig.\ \ref{fig:lux-light-quark}(b).  This model could be discovered
with a factor of 3 improvement in direct detection sensitivity, via
Higgs exchange interaction.

The sensitivity of tests from flavor-changing particle physics
processes generally depends upon different parameters than the
astrophysical ones considered above.  For example, an observable
contribution to the $B_s^0$-$\bar B_s^0$ mixing amplitude would arise
from a choice of couplings such that $|\Lambda_{2s}\Lambda_{2b}|\sim
1$ in the $d$ model.  These couplings are unconstrained by the
previous considerations.  On the other hand it is also possible to
saturate the $B_s^0$-$\bar B_s^0$ mixing bounds using
$|\Lambda_{1d}\Lambda_{1b}|\sim 1$ in the same model, but the
constraints from direction detection shown in
fig.\ \ref{fig:lux-light-quark} rule out this possibility.  This
illustrates that there can be some interplay between the particle
physics and astrophysical constraints, but that in general there is
freedom to separate them.

The nonvanishing values of $\Lambda_{ij}$ required in models 3 and 5
do not have direct implications for meson mixing; rather they imply,
through table \ref{tab:mixing}, constraints on neighboring matrix
elements.  For example $|\Lambda_{13}|=0.8$ in the $Q$ model implies
$|\Lambda_{11}|<0.02$ to satisfy $B_d^0$-$\bar B_d^0$ mixing
constraints.  Even if $\Lambda_{11}=0$ at tree level, a one-loop
vertex correction of order
\be
	\delta \Lambda_{11} \sim 
	{g^2\,\Lambda_{13}\, V_{td}\over 32\pi^2} \sim 10^{-4}
\ee
is induced by $W^\pm$ exchange.  Therefore no fine-tuning is needed to
satisfy this constraint.  Similarly $|\Lambda_{12}|=0.04$ in the $Q$
model requires $|\Lambda_{11}| < 0.05$ to satisfy $K^0$-$\bar K^0$
mixing constraints; this is also easily compatible with the maximum
size of loop corrections.

\section{Conclusions}
\label{conclusions}

Scalar flavored DM is somewhat more strongly constrained than its
fermionic analog because of the large cross section for producing the
colored fermionic mediators of SFDM at the LHC: they must typically
be at the TeV scale or higher, with the possibility for lower masses
$m_\chi \gtrsim 500$ GeV only if the dark matter is heavy, $m_\phi
\sim 300$ GeV.  As a result, many flavor-changing processes are
suppressed even without any hierarchical or MFV structure in the new
flavor-violating Yukawa couplings.  Moreover annihilation processes
relying upon mediator exchange can only be dominant for heavy DM,
$m_\phi > 300$ GeV.

Also distinct from fermion FDM, scalar FDM can have important
interactions through the Higgs portal.  These tend to dominate in DM
annihilation processes, and will be generated at one loop by the
mediator couplings $\Lambda_{ij}$ even if absent at tree level.  We
showed as a result that it is unnatural to expect mediator-dominated
annihilations outside of the heavy DM mass range mentioned above.
Another novel consequence is that there can be destructive
interference between Higgs and mediator exchange for DM scattering
on nucleons, allowing for relaxation of direct detection constraints
relative to models with only one kind of interaction.  This makes it
possible for SFDM to be relevant for indirect detection by gamma
rays from the galactic center or dwarf spheroidals, unlike for minimal
scalar DM coupled through the Higgs.

Low-energy data from $\Delta F=2$ meson mixing provide some of the
most stringent constraints on the dark matter couplings, summarized in
table \ref{tab:mixing}.  The couplings $\Lambda_{ij}$ for $i,j=1,2$
must either be very small, or very close to being diagonal.  This is
in contrast to the large values $\Lambda_{i3}$ required if DM
annihilation into top quarks is significant for determining the relic
density.  

There is one intriguing exception: $B_s$-$\bar B_s$ mixing. The
measured values of the magnitude and phase of the mixing amplitude
$M_{12}$ are consistent with the SM predictions.  However, because of
large experimental errors or theoretical uncertainties, there is ample
room for a new-physics contribution to $M_{12}$. We find that, for
reasonable values of its couplings, scalar flavored DM can contribute
significantly to both $\Delta M_s$ ($=2|M_{12}|$) and the CP-violating
phase $\beta_s$ ($=\pm {\rm arg}(M_{12})$).

Another interesting possibility in the case of light dark matter is
the apparently flavor-violating top quark decay $t\to
c\,\phi_t\,\phi_c^*$, which does not rely upon any off-diagonal
couplings since flavor conservation is invisibly accomplished by the
dark matter flavors.

An important caveat to our analysis which could deserve further study
is the simultaneous presence of mediators that couple to both left-
and right-handed quarks, and which mix with each other.  By excluding
this more elaborate class of models, we found that all amplitudes
involving mediator exchange were suppressed by $1/m_\chi^2$ (typically
times a quark mass) rather than $1/m_\chi$.  But in more complicated
models with mediators coupling to both chiralities, one could expect
much larger amplitudes involving mediators, both for DM annihilation
and for FCNC processes.

\appendix

\section{Simulation of $\chi\to\phi t$ events}
\label{app}

\begin{figure}
\centering
\includegraphics[width=1\columnwidth]{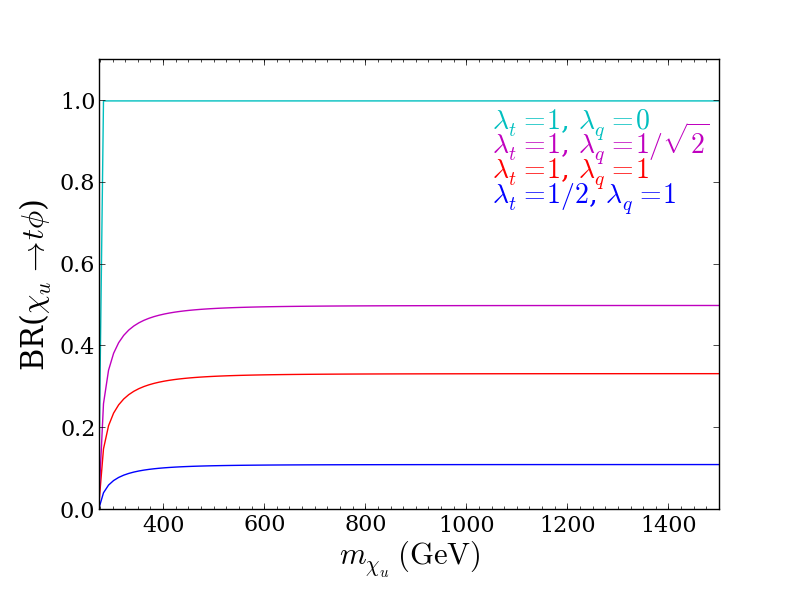}
\caption{Branching fraction of mediator decays to top quarks, $\chi_u \rightarrow t \phi$, for chosen values of quark couplings. $\lambda_q$ denotes couplings to light quarks, $\lambda_q = \lambda_u = \lambda_c$.}
\label{fig:brs}
\end{figure}

Here we give details of our simulation of the production and decay of
$u$-like mediators that decay to DM plus top quarks.  We generate
events in \textsc{MadGraph}, interfaced with \textsc{Pythia-6.4}
\cite{Sjostrand:2006za} for showering and hadronization. Events are
generated with up to one additional jet at the matrix-element level
and matched using a shower-k$_T$ scheme. Events are then passed to PGS
\cite{Conway:pgs} to simulate detector response and event
reconstruction. The anti-k$_{T}$ algorithm is used for jet
reconstruction, with jet size parameter $R=0.4$. Event selection is
performed with cuts corresponding to the signal regions defined in
ref. \cite{Aad:2014wea}. The same $K$-factor of $1.5$ is applied to
the cross section. We choose discrete values of the couplings so as to
vary the branching fraction for decays to tops. As the branching
fraction depends non-trivially on the mediator mass in the very
low-mass region, we give the branching fraction as a function of the
mediator mass in fig. \ref{fig:brs}, for the chosen values of the
couplings.  We allow for coupling to all three generations, but the
distinction between first- and second-generation quarks does not
affect the result, for a fixed value of the top branching fraction.

\bigskip
\noindent
{\bf Acknowledgments}: DL would like to thank J. Gaboriaud and A. Lenz
for input regarding the SM contribution to $B_s$-$\bar B_s$ mixing. AD
thanks Hiren Patel and Shanmuka Shivashankara for help with certain
calculations. JC thanks Mike Trott for helpful discussions,
and the Niels Bohr International Academy for its kind hospitality
during the completion of this work.
 This work was financially supported by the IPP (BB), by
NSERC of Canada (BB, DL, JMC, GD), by FQRNT of Qu\'ebec (JMC, GD)
and by the National Science
Foundation (AD) under Grant No.\ NSF PHY-1414345.

\end{document}